\documentclass[letterpaper,twocolumn,10pt]{article}
\usepackage{usenix}
\usepackage{cite}
\usepackage{todonotes}
\usepackage{acronym}
\usepackage{xspace}
\usepackage{tikz}
\usepackage{amsmath}
\usepackage{graphicx} 
\usepackage{pdflscape}
 \usepackage{multirow}
\usepackage{pifont}
\usepackage{xcolor}
\usepackage{nicematrix}
\usepackage{tabularx}
\usepackage{makecell}
\usepackage{slashbox, pict2e}
\usepackage{caption}
\usepackage{enumitem}
\usepackage{color, colortbl}
\usepackage[most]{tcolorbox}
\usepackage[verbose]{placeins}
\usepackage{balance}
\usepackage{soul}
\usepackage{etoolbox}

\newcommand{\scaledDing}[1]{\scalebox{0.75}{\ding{#1}}}

\newcounter{numFindings}
\newcounter{numGaps}
\newcommand{\finding}[1]{\begin{tcolorbox}[boxrule=0pt,frame hidden,sharp corners,enhanced,borderline west={3pt}{0pt}{violet}]\textbf{\ac{KF} \arabic{numFindings}.\stepcounter{numFindings}} #1 \end{tcolorbox}}
\newcommand{\gap}[1]{\begin{tcolorbox}[boxrule=0pt,frame hidden,sharp corners,enhanced,borderline west={3pt}{0pt}{yellow}]\textbf{\ac{RG} \arabic{numGaps}.\stepcounter{numGaps}} #1 \end{tcolorbox}}

\makeatletter
\def\hlinewd#1{%
\noalign{\ifnum0=`}\fi\hrule \@height #1 \futurelet
\reserved@a\@xhline}
\makeatother

\newtoggle{revision} 
\newtoggle{minorrev}

\newcommand{\addition}[1]{\iftoggle{revision}{{\color{blue} #1}}{#1}}
\newcommand{\minoraddition}[1]{\iftoggle{minorrev}{{\color{blue} #1}}{#1}}

\begin{document}

\date{}


\title{\Large \bf SoK: Come Together -- Unifying Security, Information Theory, and Cognition for a Mixed Reality Deception Attack Ontology \& Analysis Framework}

\author{{\rm Ali Teymourian}$^1$, {\rm Andrew M. Webb}$^1$, {\rm Taha Gharaibeh}$^{1,2}$, {\rm Arushi Ghildiyal}$^1$, {\rm Ibrahim Baggili}$^{1,2}$\\
$^1$Division of Computer Science \& Engineering\\
$^2$Baggil(i) Truth (BiT) Lab, Center for Computation and Technology\\ 
Louisiana State University\\
\{ateymo1, andrewwebb, tghara1, aghild1, ibaggili\}@lsu.edu
}


\maketitle
\thispagestyle{empty}
\pagestyle{empty}


\acrodef{MR}[MR]{Mixed Reality}
\newcommand{\MR}{\ac{MR}\xspace}

\acrodef{VR}[VR]{Virtual Reality}
\newcommand{\VR}{\ac{VR}\xspace}

\acrodef{VE}[VE]{Virtual Environment}
\newcommand{\VE}{\ac{VE}\xspace}

\acrodef{DAF}[DAF]{Deception Analysis Framework}
\newcommand{\DAF}{\ac{DAF}\xspace}

\acrodef{AR}[AR]{Augmented Reality}
\newcommand{\AR}{\ac{AR}\xspace}

\acrodef{AV}[AV]{Augmented Virtuality}
\newcommand{\AV}{\ac{AV}\xspace}

\acrodef{PMDA}[PMDA]{Perceptual Manipulation Deception Attacks}
\acrodefplural{PMDA}{Perceptual Manipulation Deception Attacks}
\newcommand{\PMDA}{\ac{PMDA}\xspace}

\acrodef{PMA}[PMA]{Perceptual Manipulation Attacks}
\acrodefplural{PMA}{Perceptual Manipulation Attacks}
\newcommand{\PMA}{\ac{PMA}\xspace}

\acrodef{SNR}[SNR]{Signal-to-Noise Ratio}
\newcommand{\SNR}{\ac{SNR}\xspace}

\acrodef{UI}[UI]{User Interface}
\newcommand{\UI}{\ac{UI}\xspace}

\acrodef{HMD}[HMD]{Head-Mounted Display}
\newcommand{\HMD}{\ac{HMD}\xspace}

\acrodef{HUD}[HUD]{Heads-Up Display}
\newcommand{\HUD}{\ac{HUD}\xspace}

\acrodef{APA}[APA]{American Psychological Association}
\newcommand{\APA}{\ac{APA}\xspace}

\acrodef{WMC}[WMC]{Working Memory Capacity}
\newcommand{\WMC}{\ac{WMC}\xspace}

\acrodef{CSI}[CSI]{Channel State Information}
\newcommand{\CSI}{\ac{CSI}\xspace}

\acrodef{RDW}[RDW]{Redirected Walking}
\newcommand{\RDW}{\ac{RDW}\xspace}

\acrodef{KF}[KF]{Key Finding}
\acrodef{RG}[RG]{Research Gap}

\acrodef{DARPA}[DARPA]{Defense Advanced Research Projects Agency}
\newcommand{\DARPA}{\ac{DARPA}\xspace}

\acrodef{CVSS}[CVSS]{Common Vulnerability Scoring System}
\newcommand{\CVSS}{\ac{CVSS}\xspace}

\begin{abstract}
We present a primary attack ontology and analysis framework for deception attacks in Mixed Reality (MR). This is achieved through multidisciplinary Systematization of Knowledge (SoK), integrating concepts from MR security, information theory, and cognition. While MR grows in popularity, it presents many cybersecurity challenges, particularly concerning deception attacks and their effects on humans. In this paper, we use the Borden-Kopp model of deception to develop a comprehensive ontology of MR deception attacks. Further, we derive two models to assess impact of MR deception attacks on information communication and decision-making. The first, an information-theoretic model, mathematically formalizes the effects of attacks on information communication. The second, a decision-making model, details the effects of attacks on interlaced cognitive processes. Using our ontology and models, we establish the MR Deception Analysis Framework (DAF) to assess the effects of MR deception attacks on information channels, perception, and attention. Our SoK uncovers five key findings for research and practice and identifies five research gaps to guide future work.
\end{abstract}

\section{Introduction}

\MR is reshaping how we perceive and interact with our physical surroundings.
In 2023, the global \MR market surged to \$4.6 billion, fueled by leading tech giants Meta, Apple, and Microsoft~\cite{mr-market}. 
\MR headsets overlay virtual information onto the real world to assist human users, such as visualizing navigational aids on sidewalks to guide pedestrians.  
Malicious actors can exploit \MR headsets to manipulate user perceptions and cause significant physical or social harm. 
For example, attackers can guide pedestrians into traffic by obstructing their view of oncoming vehicles. 

Deception attacks pose a fundamental security threat for technologies that alter human perception of the real world.
Deceptions introduce false beliefs or interpretations in a target~\cite{hyman1989psychology}. 
Illusions, central to deception, lead to false perceptions of sensory input~\cite{jastrow1900psychology}, achieved through deceit, where truthful information is hidden or false information is shown~\cite{adar2013benevolent}. 
Using \MR, attackers can affect information communication and decision-making, such as by introducing illusions (e.g., fake pedestrian crossings) or hiding essential information (e.g., navigation arrows).
Protecting \MR users is vital, yet we lack theoretical framing to describe and analyze \MR deception attacks and their effects on human cognition. 

This paper systematizes knowledge from disparate domains, introducing a framework for evaluating 
\MR deception attacks. We address the following research questions:

\begin{itemize}[labelsep=0.1cm,leftmargin=*,labelindent=1cm]
\itemsep0em 

    \item[\textbf{RQ1:}] How does existing literature categorize \MR deception attacks? 
    \item[\textbf{RQ2:}] How do we model the effects of \MR deception attacks on information communication?
    \item [\textbf{RQ3:}] How are the cognitive processes associated with decision-making affected by \MR deception attacks?
    \item [\textbf{RQ4:}] How can we systematically analyze \MR deception attacks and their effects? 
\end{itemize}

Our multi-stage methodology synthesizes knowledge from \MR security, information theory, and cognition to derive our \MR \DAF. 
First, we derived an \MR deception attack ontology from the literature. 
Then, we integrated our ontology, an information-theoretic model of communication, and a cognitive decision-making model to derive our framework. 
Our work contributes the following:

\begin{itemize}
\itemsep0em
    \item the \textbf{first in-depth investigation of deception attacks in \MR environments}; 
    \item a \textbf{deception analysis framework for assessing the effects of \MR deception attacks} on information channels and decision-making;
    \item an \textbf{ontology of \MR deception attacks};
    \item an \textbf{information-theoretic model of \MR deception attacks} that formalizes effects on communication;
    \item a \textbf{decision-making model of \MR deception attacks} that connects cognitive processes, attacks, and effects;
    \item a \textbf{literature review of deception attacks} in \MR;
    \item an \textbf{assessment of state-of-the-art \MR technical attacks} use or potential use in deception attacks.
\end{itemize}

This paper is structured as follows. 
Section \ref{sec:background} grounds our work in foundational research. 
Section \ref{sec:methodology} outlines our methodology. 
Section \ref{sec:relwork} presents a literature review of \MR deception attacks. 
Section \ref{sec:decattacks} describes an ontology of existing attacks. 
Sections \ref{sec:information-theoretic} and \ref{sec:decision-making} develop information-theoretic and decision-making models to assess how \MR deception attacks affect communication and cognition, respectively. 
Section \ref{sec:cog-framework} introduces our \MR Deception Analysis Framework. 
\addition{Section \ref{sec:discussion} discusses implications and limitations of this work.}
Section~\ref{sec:conclusion} summarizes our contributions and suggests future work.

\section{Background}
\label{sec:background}

We ground this work in foundational research on deception, information processing, decision-making, and \MR. 

\subsection{Deception}



Deception entails intentional acts to cultivate a belief in a recipient that the deceiver considers false~\cite{mitchell1986deception,zuckerman:1981}.
In order to induce false beliefs, communication is required~\cite{mitchell:1996}.
This communication may be verbal or nonverbal.
Deception can be modeled as information processing where a  
sender presents ``truthful or false information (a signal) to an opponent (the receiver) in order to gain an advantage over the opponent''~\cite{cranford:2021}.
Separate cognitive processes exist for sender (deceiver) and receiver~\cite{jenkins:2016}. 
Accounts of deception must consider how ``information sharing is dominated by unstructured communication involving natural language and a diverse collection of nonverbal cues''~\cite{jenkins:2016}.
\MR is primarily a visual medium where deceptions will often rely on nonverbal stimuli.


\addition{
Models of deception center around interpersonal communication~\cite{buller:1996,levine2014truth,gaspar:2013,gaspar:2022,kang:2022} or information transmission~\cite{borden1999information, kopp2000information, mcwhirter:2016, mccornack1992information}.
The Interpersonal Deception Theory (IDT)~\cite{buller:1996} examines deception as an interactive, reciprocal relationship where  both senders and receivers adapt their strategies in real-time. 
IDT integrates cognitive and emotional dimensions, such as arousal and suspicion, which influence deceptive behaviors and detection mechanisms during interpersonal exchanges.
Levine’s Truth-Default Theory \cite{levine2014truth} identifies cognitive biases underlying deception detection and shows that humans generally operate under a presumption of honesty.
This facilitates efficient communication but leaves individuals vulnerable to deceit.
The Emotion Deception Model~\cite{gaspar:2013,gaspar:2022} considers how both current emotions and anticipated emotions influence decisions to use deceptions during negotiations. 
McCornack \cite{mccornack1992information} models deception as manipulations of information, emphasizing how individuals exploit conversational norms to mislead others while maintaining an appearance of cooperative communication.
Borden~\cite{borden1999information} and Kopp~\cite{kopp2000information} separately proposed models of deception that are grounded in information theory. 
The Borden-Kopp model categorizes four deception strategies for manipulating a victim's perception: Degradation (conceal information), Denial (increase uncertainty), Corruption (create false belief), and Subversion (alter information processing).
We use these strategies as the basis for the foundational organization of an \MR deception attack ontology and analysis framework.}

\subsection{Information Processing}
\label{sec:information-processing}
Information processing theory emerged as a way of understanding human cognition, particularly problem-solving and decision-making, alongside advancements in computing during the 1950s and 1960s~\cite{simon1979information}. 
In this theory, computational models describe how humans acquire, process, and store information to make decisions and take actions. 
The information processing model operates in a serial manner.
First, information is input through sensory receptors in the body. 
Then, information is sequentially stored in working (short-term) memory and mentally processed in decision-making.
Finally, responses are output as human actions.
In order to avoid sensory overload, attention mechanisms filter what information is stored and processed.
We use information processing theory to derive our \MR Deception Decision-Making Model (Section~\ref{sec:decision-making}), which connects sensory input transmitted from \MR headsets to attention, memory, and other cognitive processes.

\subsection{Decision-Making}

Decision-making is a complex cognitive process that is susceptible to deception~\cite{dunbar:2014}. 
It consists of three stages~\cite{ernst2005neurobiology}. 
First, sensory input is processed to make assessments and predictions on possible outcomes.
Second, cognitive processes select an action based on the perception of inputs and predictions of outcomes.
Third, action responses are assessed to evaluate the outcome.
Individuals often do not evaluate risks based on mathematical probabilities~\cite{prospect-theory}. 
Instead, psychological factors, such as the certainty effect, play crucial roles. 
With the certainty effect, humans give more weight to outcomes that are seen as certain compared to those that are merely probable.
This insight is valuable when anticipating \MR user responses to deceptive stimuli, where the perception of risk and reward can be manipulated.
Niforatos et al.~\cite{niforatos} point out the complexities of ethical decision-making within \MR, emphasizing the impact that immersive technologies have on human cognitive evaluations.

\begin{figure*}[t!]
    \centering
    \includegraphics[width=.9\textwidth]{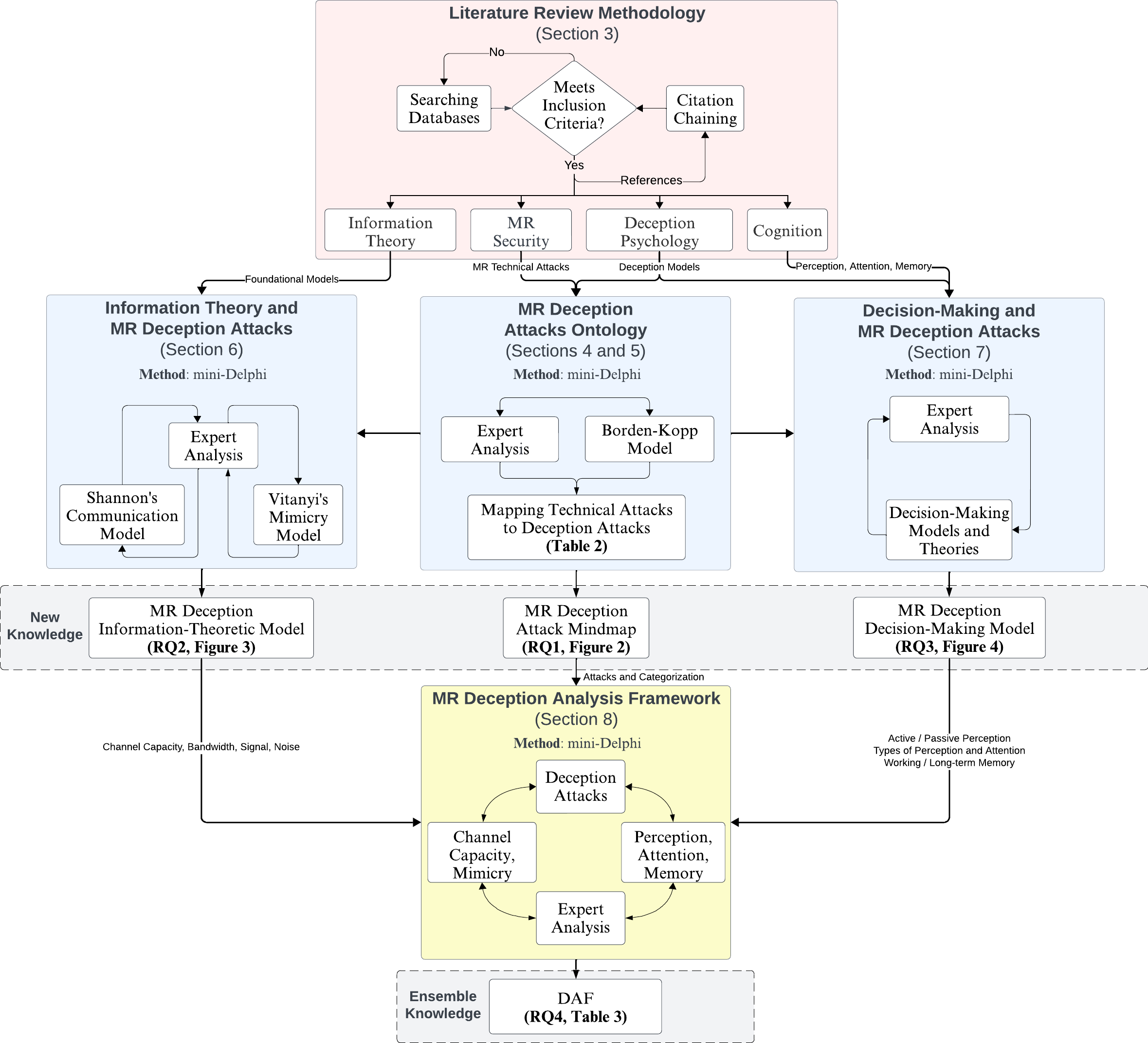}
    \caption{Our five-stage methodology beginning with literature review (top). Outcomes of the literature review informed intermediary stages. 
    Knowledge from these stages culminates in the \MR \acf{DAF}.}
    \label{fig:method}
    \vspace{-2ex}
\end{figure*}

\subsection{Mixed Reality}
Milgram and Kishino~\cite {milgram} defined \MR as a continuum of blended visual representations residing between the entirely real and the fully virtual. 
Within their continuum are two forms of \MR: \AR and \AV.
In \AR, virtual elements overlay physical reality. 
\VR headsets, such as the Apple Vision Pro or Meta Quest 3, now support \AR experiences through video pass-through where virtual information is overlaid onto camera feeds of the real world. In contrast, 
\AV integrates real-world elements into \VR. 
For example, many \VR headsets display boundaries of physical spaces as users approach to avoid collisions.
In this work, we focus on \AR and \AV systems that facilitate complex information processing scenarios. 
This complexity raises questions about how users interact with information in \MR and the potential cognitive risks or vulnerabilities in decision-making. 
To the best of our knowledge, this is the first work exploring the impact of \MR deception attacks on information communication and decision-making.






\section{Methodology}
\label{sec:methodology}

We employed a systematic methodology that included an extensive literature review and development of theoretical models to describe the \MR deception attacks.
The literature review identified relevant theories, models, attacks, and empirical outcomes, which informed each step in our process (Figure~\ref{fig:method}). 
Using our review and expert knowledge, we derived an ontology of \MR deception attacks (Figure~\ref{fig:attacks-mindmap}). 
We connected technical attacks from the literature to our ontology (Table~\ref{tab:attacks}).
Subsequently, we integrated an information-theoretic perspective to examine how deception attacks impact information communication in \MR (Figure~\ref{fig:mr-communication-model}). Next, we developed a decision-making model that describes how cognitive processes handle sensory stimuli from \MR headsets (Figure~\ref{fig:process-model}). Finally, we combined our ontology and two models to derive a framework for assessing the cognitive effects of \MR deception attacks on decision-making (Table~\ref{tab:framework}). 



\emph{\textbf{Literature Review}}\addition{~(Section~\ref{sec:relwork})}: 
We conducted a systematic literature review covering a wide range of topics, including deception, privacy, perceptual manipulation, cognition, and decision-making.
We used Google Scholar, ACM Digital Library, IEEE Xplore, MIT Press, and APA PsycArticles. 
Search terms included mixes of ``\AR/\VR Security'', ``\MR Deception'', ``Perceptual Manipulation'', and ``Decision-Making''. 
We collected articles from reputable journals and conference proceedings, including  USENIX Security, S\&P, ISMAR, IEEE VR, and the Journal of Experimental Psychology. 
We limited articles to those published in the past five years to ensure relevance to current \MR technologies. 
However, we additionally took into account important historical works that had significant impact. 
We focused on articles 
with well-defined research questions, comprehensive analysis, and innovative contributions.

The filtering process began by retrieving over 200 articles from search engines and databases. Two researchers screened titles and abstracts for relevance to ensure a consensus-based approach. The criteria for relevance included: alignment with \textbf{\MR security}, \textbf{information theory}, \textbf{cognition}, and \textbf{deception psychology}; presence of well-defined research questions; and contributions to the field's innovation and depth of analysis. 
We reviewed full texts to confirm suitability based on the depth of analysis, innovation, and relevance to our research objectives. 
Articles were excluded if they lacked depth of analysis, innovation, or relevance to the key themes. 
This resulted in a final selection of ($n=80$) articles across different domains, which are categorized in Table \ref{tab:PaperCategories}.


\bgroup

\begin{table}[t!]
\caption{Reviewed Articles Classified by Research Area.} 
\vspace{-1ex}
\label{tab:PaperCategories}
\renewcommand{\arraystretch}{1}
\fontsize{9}{10}\selectfont
\begin{tabular}{|p{2.5cm}|p{4cm}|c|}
\hline
\textbf{Categories} & \textbf{Articles} & \textbf{\#}\\
\hlinewd{1.5pt}
\multicolumn{1}{|c|}{\textbf{MR Security}} & &\\ 
\rowcolor{gray!10} \textit{User Manipulation \& Deception} & \hspace{-0.1mm}\cite{Casey_2021,chandio2024stealthy,Brinkman,deHaas2022audio,wang2023dark,Trivers2011Deceit,Tseng_2022,Ali_Mahmood_Qadri_2018,cheng2023exploring,yang2024inception,Lee2021AdCube,tu2018injected,nilsson201815,ledoux2013using} & 14 \\ 
 \textit{Privacy and Data Security} & \hspace{-0.1mm}\cite{ling2019know,vondravcek2023rise,Yarramreddy,Casey,nair2023exploring,slocum2023going,al2021vr,su2024remote,odeleye2021detecting,Roesner2011Security,10.1145/2736277.2741657,10.1145/2873587.2873595,inproceedings3,shi2021face,zhang2023s,farrukh2023locin,luo2024eavesdropping, cheng2024user, Slocum2024Shared} & 19 \\ 
\rowcolor{gray!10} \textit{Frameworks / Surveys} & \hspace{-0.1mm}\cite{article1,garrido2023sok,10.3389/fict.2019.00005,De_Guzman_2019,219386, stephenson2022sok} & 6 \\ \hline 

\multicolumn{1}{|c|}{\textbf{Information Theory}} & \hspace{-0.1mm}\cite{shannon1948mathematical,Vitanyi,brumley:2012,kopp:2018,buller:1996} & 5 \\ \hline

\multicolumn{1}{|c|}{\textbf{Cognition}} & & \\
\rowcolor{gray!10} \textit{Perception} & \hspace{-0.1mm}\cite{qiong2017brief,wang2007cognitive,gibson2014ecological,Rock1983-ROCTLO,Dionisio2001Differentiation,Cetnarski2014Subliminal} & 6 \\
\textit{Attention} & \hspace{-0.1mm}\cite{James1890,posner1980orienting,posner1990attention,stevens2012role,murphy2016twenty,spelke1976skills,Herbranson2017,mackworth1948breakdown} & 8 \\ 
\rowcolor{gray!10} \textit{Memory} & \hspace{-0.1mm}\cite{alan1992working,baddeley1974working,baddeley2007working,ATKINSON196889,downing2000interactions,santangelo2013contribution,unsworth2020working,engle2002working} & 8 \\\hline

\multicolumn{1}{|c|}{\textbf{Deception Psych.}} & \hspace{-0.1mm}\cite{mitchell:1996,mitchell1986deception,hyman1989psychology,gombos:2006,jenkins:2016,cranford:2021,ekman1969repertoire,tversky1974judgment,depaulo1996lying, gaspar:2013, kang:2022, gaspar:2022, levine:2022, mcwhirter:2016} & 14 \\\hline
\multicolumn{1}{r}{} & \multicolumn{1}{|r|}{$\Sigma$ Total:} & 80 \\\cline{2-3}
\end{tabular}
\vspace{-3ex}

\end{table}
\egroup



 

\emph{\textbf{\MR Deception Attacks Ontology}}\addition{~(Section~\ref{sec:decattacks})}: After our literature review, two researchers iteratively outlined an encyclopedic map of identified deception attacks. The iterative process was enhanced using the mini-Delphi method \cite{dalkey1963experimental, pan1996mini} in which two subject matter experts reviewed and further refined the ontology across each iteration. The outcome was a mind map of deception attacks in \ac{MR} (Figure~\ref{fig:attacks-mindmap}).
Then, we characterized how technical attacks identified in our literature review fit within the newly developed ontology~(Table~\ref{tab:attacks}).

\emph{\textbf{Information Theory and Deception Attacks}}\addition{~(Section~\ref{sec:information-theoretic})}: We derive an information-theoretic model to describe how \MR deception attacks affect information communication. We employ Borden-Kopp's deception model \cite{kopp:2018} that uses Shannon's information theory \cite{shannon1948mathematical} to formulate how information is transmitted from a source (e.g., \MR application) to a user via a \MR headset. Additionally, we utilize Vitanyi's model of mimicry \cite{Vitanyi} to mathematically assess differences between source-generated messages and those created by an attacker. 

 \emph{\textbf{Decision-Making and Deception Attacks}}\addition{~(Section~\ref{sec:decision-making})}: We used our \MR Deception Ontology and our review of deception psychology and cognition literature to develop an \MR Decision-Making Model. This model connects cognitive processes of perception, attention, memory, and decision-making to types of deception attacks. This process involved a mini-Delphi approach in which renditions of the model were iteratively revised using expert knowledge and prior literature. 

 \emph{\textbf{\MR \acf{DAF}}}\addition{~(Section~\ref{sec:cog-framework})}: We utilize our two models to develop a framework for analyzing the effects of \MR deception attacks on information communication and human cognition. This framework allows for qualitatively evaluating the impact of MR deception attacks on cognitive processes associated with perception, attention, memory. 

\finding{Our multidisciplinary methodology shows how to connect disparate knowledge into an ensemble framework. As computing becomes more ubiquitous, security challenges require broader perspectives and analysis, particularly in terms of human cognition. We are not aware of other work that connects literature and theories from cybersecurity, \MR, and cognitive sciences into a cohesive framework.}




\section{\MR Attacks and Surveys}
\label{sec:relwork}
Our literature review categorizes unique aspects of \MR security into three distinct areas: User Manipulation and Deception, Privacy and Data Security, and Frameworks and Surveys.

\subsection{User Manipulation and Deception}

Prior work explored techniques to manipulate facets of users' perceptions and decision-making in \MR. 
Casey et al. \cite{Casey_2021} introduced new proof-of-concept attacks that pose a threat to user safety in a \VE. Their work categorized and defined the following attack types: chaperone, disorientation, human joystick, and overlay. The human joystick successfully manipulated users to move to specific physical locations without their awareness. The chaperone attacks manipulated the \VE boundaries, while the disorientation attack elicited a sense of dizziness and confusion from an immersed \VR user. Lastly, in an overlay attack, an adversary overlaid objects such as images and videos onto a user’s \VR view.
Chandio et al.~\cite{chandio2024stealthy} introduced stealthy and practical multi-modal attacks on \MR tracking, showing that \MR systems relying on sensor fusion algorithms for tracking can be compromised through perceptual manipulation by attacking multiple sensing streams simultaneously.

Nilsson et al. \cite{nilsson201815} provided an overview of \RDW techniques in \VR that use subtle manipulations of gains and overt redirection techniques to manipulate user's perception of space and movement. Brinkman \cite{Brinkman} describes attacks that subtly influence user choices without their awareness as decisional interference.
De Haas \& Lee \cite{deHaas2022audio} provide a comprehensive analysis of the manipulative potential of audio effects design in \AR which systematically categorizes deceptive audio cues into various categories, each of which uniquely influences user perception and behavior.
Wang et al. \cite{wang2023dark} further investigate how these deceptive design techniques, known as dark patterns, can manipulate users in \AR environments and compromise their information and safety.
Building on psychological aspects of manipulation, Trivers \cite{Trivers2011Deceit} describes how deception is a natural part of life, not just for humans but all living beings.
This analysis provides a foundational understanding of the psychological dynamics at play, illustrating how \MR systems can exploit the natural tendencies of humans to manipulate and be manipulated, influencing user perception and decision-making.

\acp{PMA} attempt to exploit a user's sensory perceptions to influence their decision-making, which can lead to physical harm \cite{Tseng_2022, cheng2023exploring}.
Ali et al.~\cite{Ali_Mahmood_Qadri_2018} investigated visual deception by creating illusions of 3D views using projections onto 2D surfaces.
Cheng et al.~\cite{cheng2023exploring} derived a framework for comprehending and addressing \acp{PMA} within the context of \MR. 
They demonstrated that \acp{PMA} can manipulate user perceptions to affect reaction times. 
They investigated the effects of \acp{PMA} on situational awareness, revealing how \MR content can divert users' attention away from essential real-world stimuli, undermining their concentration and attentiveness. 
Ledoux et al.~\cite{ledoux2013using} found that visual cues in \VR can evoke food cravings, showing how sensory manipulations influence user perception. 
Tseng et al.~\cite{Tseng_2022} investigated the risks of perceptual manipulations in \VR, focusing on the negative impacts that these manipulations may have on users.

\gap{Most recent research on user manipulation and deception has focused on \VR systems, leaving \AR systems underrepresented. Future research should prioritize \AR security.}


\subsection{Privacy and Security}
\addition{MR and VR headsets pose significant challenges for privacy and security.
These headsets collect, use, and present personal information, making them vulnerable to information leaks via side-channel attacks.
Further, attackers can use deception attacks to disrupt information channels and cognitive processes causing users to take actions that may expose additional personal information.}

\addition{
Slocum et al.~\cite{slocum2023going} introduced TyPose, which uses machine learning techniques to classify motion signals from \MR headsets by analyzing subtle head movements made by users when interacting with virtual keyboards. 
Al Arafat et al.~\cite{al2021vr} presented the VR-Spy system, which utilizes the channel state information of Wi-Fi signals to detect and recognize keystrokes based on fine-grained hand movements. Su et al.~\cite{su2024remote} present a method for remotely extracting motion data from network packets and correlating them with keystroke entries to obtain user-typed data such as passwords or private conversations.} 
Ling et al. \cite{ling2019know} highlighted the vulnerability of \VR systems to novel side-channel attacks. 
They showed how these attacks exploit computer vision and motion sensor data to infer keystrokes in a \VE. 
\addition{Knowing what information a user is typing or specific personal details could help attackers develop more believable deception attacks.}

Vondráček et al.~\cite{vondravcek2023rise} introduced the Man-in-the-Room attack in \VR, where an attacker gains unauthorized access to a private \VR room and observes all interactions. 
\addition{Through observation, attackers can develop more targeted deception attacks.} 
Nair et al. \cite{nair2023exploring} outlined significant privacy risks in \VR environments, proposing a threat model with four adversaries: Hardware, Client, Server, and User. These adversaries have access to different aspects of the \VR information flow. These risks can covertly reveal personal data, and adversarially designed \VR games may manipulate users into disclosing sensitive information.

\addition{Prior work has also explored the digital forensics of VR headsets.}
Yarramreddy et al.~\cite{Yarramreddy} presented an exploration of the forensics of immersive \VR systems, which demonstrates the feasibility of reconstructing forensically relevant data from both network traffic and the systems themselves. 
Casey et al.~\cite{Casey} introduced the first open-source \VR memory forensics plugin for the Volatility Framework. 
\addition{Using forensic techniques could allow an attacker to uncover personal information about a user's behavior or interest, which could be leveraged for deception attacks.}

\addition{
Security issues can expose \MR users to physical harm and potential deception attacks.}
Odeleye et al.~\cite{odeleye2021detecting} showed attacks targeting GPU and network vulnerabilities in \VR systems to manipulate frame rates and cause \VR sickness. 
Roesner et al.~\cite{Roesner2011Security} conducted a comprehensive examination of security and privacy concerns in \AR, unveiling new vulnerabilities unique to \AR applications. 
\addition{For example, they suggest displaying the provenance of AR elements so that users know the source of augmentations. Without this, users are susceptible to deception attacks that inject false information.}
McPherson et al.~\cite{10.1145/2736277.2741657} conducted the first system-level assessment of security and privacy features in \AR browsers. 
Lebeck et al.~\cite{10.1145/2873587.2873595} introduced Arya, an \AR platform aimed at regulating application output to mitigate risks from malicious or faulty applications. 
This focus on output security is complemented by research delving into input privacy risks and the largely unexplored area of malicious \AR output \cite{inproceedings3}.
Cheng et al. \cite{cheng2024user} introduced several proof-of-concept attacks targeting \UI security vulnerabilities in \AR systems. 
Slocum et al. \cite{Slocum2024Shared} investigate the security vulnerabilities in multi-user \AR applications, focusing on the shared state that maintains a consistent virtual environment across users.


\begin{figure*}[t!]
    \centering
    \includegraphics[width=\textwidth]{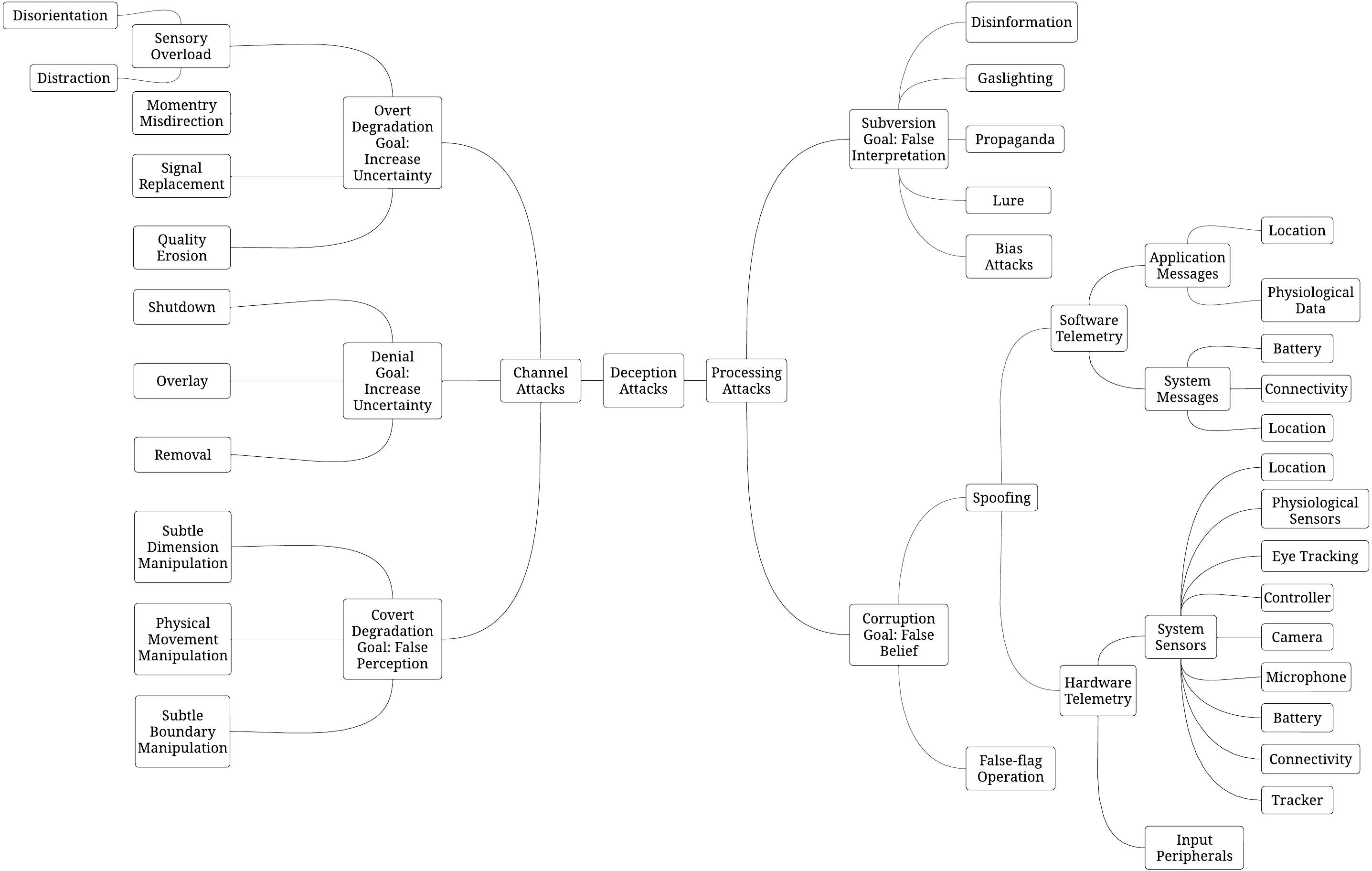}
    \caption{Mind Map of \MR Deception Attacks Ontology. Channel attacks on the left. Processing attacks on the right.} 
    \label{fig:attacks-mindmap}
    \vspace{-1ex}
\end{figure*}

\subsection{Frameworks and Surveys} 

Garrido et al.~\cite{garrido2023sok} systematized knowledge on \VR privacy threats and countermeasures, 
focusing on two types of attacks: profiling and identification.
Profiling attacks collect sensitive personal data to create user profiles.
Identification attacks uniquely pinpoint a user within a \VR environment. 
Happa et al.~\cite{10.3389/fict.2019.00005} developed an abstraction-based reasoning framework to reveal possible attacks in collaborative \MR applications.
De Guzman et al. \cite{De_Guzman_2019} provided a survey of various protection mechanisms proposed for \MR. 
Adams et al.~\cite{219386} conducted interviews with \MR users and developers to survey \MR privacy policies and their perceptions. 
Stephenson et al. \cite{stephenson2022sok} systematized knowledge on \AR/\VR authentication mechanisms, evaluating research proposals and practical deployments.

\gap{There is a notable lack of frameworks that address diverse aspects of \MR security, including technical exploits, user experience, detection, and defense.}

\section{\MR Deception Attacks Ontology}
\label{sec:decattacks}

We derive a \MR deception attack ontology (Figure~\ref{fig:attacks-mindmap}) from our review of the literature, our expert knowledge, and the Borden-Kopp model~\cite{borden1999information,kopp2000information,brumley:2012}.  
\addition{Their model focuses on how deceptions alter a victim's decision-making by manipulating information channels to inject false information or hide true information.
Additionally, they identify how false information can be used to induce biases that influence how information is processed and interpreted. 
As MR headsets directly transmit information to users, their information-theoretic model provides an appropriate and robust framework for describing, categorizing, and analyzing MR deception attacks.
}
The Borden-Kopp model divides deception attacks into \textit{channel attacks} and \textit{processing attacks}. 



\subsection{Channel Attacks}
Channel attacks primarily target information communication channels. 
These attacks exploit the physical or logical paths that data takes as it moves between different components of a system or between different entities. 
The Borden-Kopp model identifies three types of channel attacks: \textit{overt degradation}, \textit{covert degradation}, and \textit{denial}.

\subsubsection{Overt Degradation} 
     With overt degradation, attackers of \MR systems create confusion by introducing substantial visual, auditory, or tactile noise to prevent victims from accurately perceiving or engaging with virtual objects, the physical world, or associated tasks.
     Due to the blatant nature of overt degradation, victims become aware that they are under attack.
     The presence of virtual noise can be disorienting in the context of \MR, as users heavily depend on the seamless integration of real and virtual information in order to maintain focus on a task. 
     \addition{Further, it can disrupt immersive experiences, preventing users from becoming fully engaged in a task.}
     We identify the following forms of overt degradation attacks:
     
     \begin{itemize}
     \itemsep0em
     \item 
     \emph{Sensory Overload}: Inundate the user's sensory receptors with excessive amounts of stimuli, leading to disorientation or distraction \cite{Roesner2011Security, odeleye2021detecting}. 
     Disorientation can cause a user to feel lost or confused within a \VE, making them more susceptible to manipulation. 
     Distraction diverts the user's attention, potentially preventing them from detecting or responding to an attack.
     
     \item \emph{Momentary Misdirection}: Redirect the user's attention using virtual content within a \MR systems. Misdirection distracts the user from their task. For example, an attacker can insert flashing virtual elements that draw the user's visual attention away from seeing important information or activities in the physical world.
     
     \item  \emph{Signal Replacement}: Alter or replace sensory input within \MR systems. This can lead to a user perceiving a different reality from what actually exists, potentially causing confusion, disorientation, or exploitation \cite{Tseng_2022}.
     
     \item  \emph{Quality Erosion}: Reduce the quality of the signal from the \MR headset. This can be achieved through actions such as decreasing the resolution of visual elements, introducing distortions to audio, or reducing the vibration intensity of haptic feedback.
     \end{itemize}

\subsubsection{Covert Degradation}
      Covert degradation attacks subtly suppress or diminish the clarity of information presented by \MR headsets.
      Attackers can blend deceptions seamlessly with the \MR environment, thereby making it harder for the user to discern.
      \addition{By leveraging immersive MR experiences, deception attacks can mask false information as users' attention and interactions are focused elsewhere.}
      We identify the following forms of covert degradation attacks:
      
     \begin{itemize}
     \itemsep0em
     \item \emph{Physical Movement Manipulation}: Relocate a user without their awareness or consent by discreetly shifting the center of a \VE while they focus on a task
     ~\cite{Casey_2021}.
     
     \item \emph{Boundary Manipulation}: Altering boundaries within the \VE, which can lead to unexpected collisions with objects or distortions in spatial perception~\cite{schmidt2019blended}.
     
     \item \emph{Dimension Manipulation}: Modifying the proportions, scale, or spatial relationships of virtual objects~\cite{bozgeyikli2021evaluating}.
     \end{itemize} 

\subsubsection{Denial}

Denial attacks seek to increase uncertainty by obstructing the user's access to information. This is achieved by shutting down virtual overlays, prohibiting interaction with virtual objects, or disrupting the seamless blend of real and virtual elements. 
This is often an overt method of deception, as users may be cognizant of their deprived or diminished accesses \cite{kopp2003shannon}. A user may find themselves subject to a Denial attack if they lose ingress to existing networks, communication channels, and various other system features. 
We identify the following forms of denial attacks:
    \begin{itemize}
    \itemsep0em 
    
    \item \emph{Shutdown}: Deliberately terminate or disable a \MR communication channel or service. 
    
    \item \emph{Overlay}: Layer content over a communication channel to disrupt normal operations of the channel \cite{Lee2021AdCube, Roesner2011Security, Tseng_2022}. 
     
    \item \emph{Removal}: Selectively remove or block information \cite{odeleye2021detecting}.
    \end{itemize} 

\bgroup
\begin{table*}[t!]
\caption{Connecting Technical Attacks to \MR Deception Attacks.} 
\vspace{-2ex}
\fontsize{9}{10}\selectfont
\centering
\renewcommand{\arraystretch}{1.25}
\setlength{\tabcolsep}{3pt}

\begin{NiceTabular}{c|ccc|ccccc|cccc|ccc|ccc|cc|ccccc|}
\cline{2-26}
& \multicolumn{3}{p{1cm}|}{\multirow{2}{1cm}{\centering Sensory Modality}} & \multicolumn{5}{c|}{\multirow{2}{1cm}{\centering Technical Modality}} & \multicolumn{10}{c|}{Channel Attacks} & \multicolumn{7}{c|}{Processing Attacks} \\
\cline{10-26} 
& \multicolumn{3}{c|}{} & \multicolumn{5}{c|}{} & \multicolumn{4}{c|}{OD} & \multicolumn{3}{c|}{CD} & \multicolumn{3}{c|}{Denial} & \multicolumn{2}{c|}{Corr.} & \multicolumn{5}{c|}{Subversion} \\
\multirow{-3}{*}{  
} 
\textbf{Technical Attacks} & \rotatebox{90}{\textbf{Visual}} & \rotatebox{90}{\textbf{Auditory}} & \rotatebox{90}{\textbf{Tactile}} & \rotatebox{90}{\textbf{Hardware}} & \rotatebox{90}{\textbf{Software}} & \rotatebox{90}{\textbf{Network}} & \rotatebox{90}{\textbf{Data}} &  \rotatebox{90}{\textbf{Side-Channel}} & {{\rotatebox{90}{\textbf{Sensory Overload}}}} & {{\rotatebox{90}{\textbf{Momentary Misdirection}}}} & {{\rotatebox{90}{\textbf{Signal Replacement}}}} & {{\rotatebox{90}{\textbf{Quality Erosion}}}} & {{\rotatebox{90}{\textbf{Subtle Boundary Manip.}}}} & {{\rotatebox{90}{\textbf{Subtle Dimension Manip.}}}} & {{\rotatebox{90}{\textbf{Physical Movement Manip.}}}} & \rotatebox{90}{\textbf{Shutdown}} & \rotatebox{90}{\textbf{Overlay}} & \rotatebox{90}{\textbf{Removal}} & {{\rotatebox{90}{\textbf{Spoofing}}}} & {{\rotatebox{90}{\textbf{False-Flag Operations}}}} & \rotatebox{90}{\textbf{Bias Attacks}} & \rotatebox{90}{\textbf{Lure}} & \rotatebox{90}{\textbf{Disinformation}} & \rotatebox{90}{\textbf{Propaganda}} & \rotatebox{90}{\textbf{Gaslighting}} \\
\hlinewd{1.5pt}
\rowcolor{gray!10}
\multicolumn{1}{|r|}{GPU-Based \cite{odeleye2021detecting}} & {\ding{51}} &  & &  & {\scaledDing{110}} &  &  &  & {\scaledDing{109}} &  &  & {\scaledDing{108}} &  &  &  & {\scaledDing{108}} &  &  &  &  &  &  &  & &\\

\multicolumn{1}{|r|}{Network-Based \cite{odeleye2021detecting}} & {\ding{51}} & {\ding{51}} & {\ding{51}} &  &  & {\scaledDing{110}} &  &  &  &  &  & {\scaledDing{108}} &  &  &  & {\scaledDing{109}} &  &  &  &  &  &  &  & &\\

\rowcolor{gray!10}
\multicolumn{1}{|r|}{Color \cite{cheng2023exploring}} & {\ding{51}} &  &  &  & {\scaledDing{110}} &  &  &  &  &  &  &  &  &  &  &  & {\scaledDing{108}} &  &  &  &  &  &  & &\\

\multicolumn{1}{|r|}{Auditory \cite{cheng2023exploring}} &  & {\ding{51}} &  &  & {\scaledDing{110}} &  &  &  & {\scaledDing{108}} & {\scaledDing{108}} &  &  &  &  &  &  &  &  &  &  &  &  &  & &\\

\rowcolor{gray!10}
\multicolumn{1}{|r|}{Puppetry \cite{Tseng_2022}} & {\ding{51}} &  & \ding{51} &  & {\scaledDing{110}} &  & {\scaledDing{111}} &  &  &  &  &  &  &  & {\scaledDing{108}} &  &  &  &  &  &  & {\scaledDing{108}} & & & \\

\multicolumn{1}{|r|}{Mismatching \cite{Tseng_2022}} & {\ding{51}} &  &  &  & {\scaledDing{110}} &  & {\scaledDing{111}} &  &  &  &  &  &  & {\scaledDing{108}} &  &  & {\scaledDing{108}} & {\scaledDing{108}} &  &  &  &  & {\scaledDing{109}} & & {\scaledDing{109}}\\

\rowcolor{gray!10}
\multicolumn{1}{|r|}{Object-in-the-Middle \cite{cheng2024user}} & {\ding{51}} &  &  &  & {\scaledDing{110}} &  &  &  &  &  &  &  &  & {\scaledDing{109}} &  &  & {\scaledDing{108}} &  &  &  &  &  &  & & \\

\multicolumn{1}{|r|}{Object Erasure \cite{cheng2024user}} & {\ding{51}} &  &  &  & {\scaledDing{110}} &  &  &  &  &  &  &  & {\scaledDing{109}} & {\scaledDing{109}} &  &  &  & {\scaledDing{108}} &  &  &  &  &  & & \\

\rowcolor{gray!10}
\multicolumn{1}{|r|}{Chaperone \cite{Casey_2021}} & {\ding{51}} &  &  &  & {\scaledDing{110}} & {\scaledDing{111}} &  &  &  &  &  &  & {\scaledDing{108}} &  &  &  &  &  &  &  &  & {\scaledDing{109}} & {\scaledDing{109}} & &\\

\multicolumn{1}{|r|}{Human-Joystick \cite{Casey_2021}} & {\ding{51}} &  &  &  & {\scaledDing{110}} & {\scaledDing{110}} & {\scaledDing{110}} &  &  &  &  &  &  &  & {\scaledDing{108}} &  &  &  &  &  &  & {\scaledDing{109}} &  & &\\

\rowcolor{gray!10}
\multicolumn{1}{|r|}{Inception \cite{yang2024inception}} & {\ding{51}} &  &  &  & {\scaledDing{110}} & {\scaledDing{111}} & {\scaledDing{110}} &  &  &  &  &  &  &  & {\scaledDing{108}} &  &  &  & {\scaledDing{109}} &  &  &  & {\scaledDing{109}} & &\\

\multicolumn{1}{|r|}{Man in the Room \cite{vondravcek2023rise}} & {\ding{51}} &  &  &  & {\scaledDing{110}} &  {\scaledDing{110}} & {\scaledDing{110}} & {\scaledDing{111}} &  &  &  &  &  &  &  &  &  &  & \scaledDing{109} &  &  &  & \scaledDing{109} & & \scaledDing{109}\\

\rowcolor{gray!10}
\multicolumn{1}{|r|}{Output Manipulation \cite{Roesner2011Security}} & {\ding{51}} &  &  &  & {\scaledDing{110}} &  &  &  &  &  &  &  &  &  &  &  & {\scaledDing{108}} &  &  &  &  &  & {\scaledDing{109}} & &\\

\multicolumn{1}{|r|}{Clickjacking \cite{Roesner2011Security}} & {\ding{51}} &  &  &  & {\scaledDing{110}} &  &  &  &  & {\scaledDing{109}} &  &  &  &  &  &  & {\scaledDing{108}} &  &  &  &  &  &  & &\\

\rowcolor{gray!10}
\multicolumn{1}{|r|}{Cursor-Jacking \cite{Lee2021AdCube}} & {\ding{51}} &  &  &  & {\scaledDing{110}} &  &  &  &  &  & {\scaledDing{108}} &  &  &  &  &  & {\scaledDing{108}} &  &  &  &  &  &  & &\\

\multicolumn{1}{|r|}{Blind Spot \cite{Lee2021AdCube}} & {\ding{51}} &  &  &  & {\scaledDing{110}} &  &  &  &  &  & {\scaledDing{109}} &  & {\scaledDing{108}} &  &  &  & {\scaledDing{108}} &  &  &  &  &  &  & &\\

\rowcolor{gray!10}
\multicolumn{1}{|r|}{Read \cite{Slocum2024Shared}} & {\ding{51}} &  &  &  & {\scaledDing{110}} &  & {\scaledDing{110}} &  &  &  & {\scaledDing{108}} &  &  &  &  &  & {\scaledDing{108}} &  &  &  &  &  & {\scaledDing{109}} & &\\

\multicolumn{1}{|r|}{Write \cite{Slocum2024Shared}} & {\ding{51}} &  &  &  & {\scaledDing{110}} &  & {\scaledDing{110}} &  &  &  & {\scaledDing{108}} &  &  &  &  &  & {\scaledDing{108}} &  & {\scaledDing{108}} & {\scaledDing{109}} &  &  &  & &\\

\rowcolor{gray!10}
\multicolumn{1}{|r|}{Hand Gestures Inference \cite{zhang2023s}} & {\ding{51}} &  &  &  &  &  &  & {\scaledDing{110}} &  & {\scaledDing{109}} &  &  &  &  &  &  & {\scaledDing{108}} &  &  &  &  &  &  & &\\

\multicolumn{1}{|r|}{Face-Mic \cite{shi2021face}} &  & {\ding{51}} &  & {\scaledDing{111}} &  &  &  & {\scaledDing{110}} &  &  & {\scaledDing{109}} &  &  &  &  &  &  &  & {\scaledDing{109}} & {\scaledDing{109}} &  &  &  & &\\

\rowcolor{gray!10}
\multicolumn{1}{|r|}{TyPose \cite{slocum2023going}} &  &  & {\ding{51}} &  &  &  &  & {\scaledDing{110}} &  &  & {\scaledDing{109}} &  &  &  &  &  &  &  & {\scaledDing{109}} & {\scaledDing{109}} &  &  &  & &\\

\multicolumn{1}{|r|}{VRSpy \cite{al2021vr}} & {\ding{51}} &  &  &  &  &  &  & {\scaledDing{110}} &  &  & {\scaledDing{109}} &  &  &  &  &  &  &  & {\scaledDing{109}} & {\scaledDing{109}} &  &  &  & &\\

\rowcolor{gray!10}
\multicolumn{1}{|r|}{Remote Keylogging \cite{su2024remote}} & {\ding{51}} &  &  &  &  &{\scaledDing{111}}  &  & {\scaledDing{110}} &  &  &  &  &  &  &  &  &  &  & {\scaledDing{109}} & &  &  &  & &\\

\multicolumn{1}{|r|}{Heimdall \cite{luo2024eavesdropping}} &  & {\ding{51}} &  &  &  &  &  & {\scaledDing{110}} &  &  & {\scaledDing{109}} &  &  &  &  &  &  &  & {\scaledDing{109}} & &  &  &  & &\\

\rowcolor{gray!10}
\multicolumn{1}{|r|}{LocIn \cite{farrukh2023locin}} & {\ding{51}} &  &  &  & {\scaledDing{110}} &  & {\scaledDing{110}} &  &  &  &  &  &  &  & {\scaledDing{109}} &  &  &  & {\scaledDing{109}}  &  &  &  &  &  &\\

\multicolumn{1}{|r|}{Zero Displacement \cite{chandio2024stealthy}} & {\ding{51}} &  &  & {\scaledDing{110}} & {\scaledDing{110}} &  & {\scaledDing{110}} & {\scaledDing{110}} &  & {\scaledDing{109}} &  &  &  & {\scaledDing{108}} &  &  &  &  &  &  &  &  & {\scaledDing{109}} & &\\

\rowcolor{gray!10}
\multicolumn{1}{|r|}{Speed Manipulation \cite{chandio2024stealthy}} & {\ding{51}} &  &  & {\scaledDing{110}} & {\scaledDing{110}} &  & {\scaledDing{110}} & {\scaledDing{110}} &  &  &  &  &  &  & {\scaledDing{108}} &  &  &  & {\scaledDing{108}} &  & {\scaledDing{109}} &  &  & &\\

\multicolumn{1}{|r|}{Path Deviation \cite{chandio2024stealthy}} & {\ding{51}} &  &  & {\scaledDing{110}} & {\scaledDing{110}} &  & {\scaledDing{110}} & {\scaledDing{110}} &  &  & {\scaledDing{108}} &  & {\scaledDing{108}} &  &  &  &  &  & {\scaledDing{108}} &  &  &  & {\scaledDing{109}} & &\\

\rowcolor{gray!10}
\multicolumn{1}{|r|}{Side-Swing and Switching \cite{tu2018injected}} & {\ding{51}} & {\ding{51}} &  & {\scaledDing{110}} &  &  &  & {\scaledDing{110}} &  &  &  &  &  &  &  &  &  &  & {\scaledDing{108}} &  & {\scaledDing{108}} &  &  & &\\

\multicolumn{1}{|r|}{Fabrication of False Narratives \cite{brown2023misinformation}} & {\ding{51}} &  &  &  & {\scaledDing{110}} &  &  &  &  &  &  &  &  &  &  &  &  &  & {\scaledDing{108}} &  &  &  & {\scaledDing{108}} & &\\

\rowcolor{gray!10}
\multicolumn{1}{|r|}{Non-Verbal Manipulative Persuasion \cite{brown2023misinformation}} & {\ding{51}} & {\ding{51}} &  &  & {\scaledDing{110}} &  &  &  &  &  & {\scaledDing{108}} &  &  &  &  &  &  &  &  &  &  &  &  & {\scaledDing{108}} &\\

\multicolumn{1}{|r|}{Selective Exposure \cite{brown2023misinformation}} & {\ding{51}} & {\ding{51}} &  &  & {\scaledDing{110}} &  &  &  &  &  & {\scaledDing{108}} &  &  &  &  &  &  &  &  &  & {\scaledDing{108}} &  &  &  &\\

\hline
\end{NiceTabular}

\fontsize{10}{11}\selectfont
\vspace{0.5em}
\scaledDing{110} Primary Technical Modality \scaledDing{111} Secondary Technical Modality\\
\scaledDing{108} Mentioned in the article \scaledDing{109} Not specifically mentioned, but can be deployed using the attack
\vspace{-1ex}

\label{tab:attacks}
\end{table*}
\egroup

\subsection{Processing Attacks}
Processing attacks target vulnerabilities in how humans cognitively process information, aiming to deceive humans by altering their perceptions, interpretations, and understandings of information. 
The Borden-Kopp model identifies two types of processing attacks: Corruption and Subversion.

    \subsubsection{Corruption} 
Corruption attacks deliberately manipulate the \MR system by counterfeiting existing virtual elements and information.
These manipulations result in inconspicuous data and actions that are difficult to discern from standard data and actions within the \MR system.
Their primary objective is to create false belief in a user, often causing compromised decision-making, incorrect conclusions, or virtual misdirection. 
\addition{Due to the immersiveness of MR, users may be more susceptible to corruption attacks as their engagement keeps them preoccupied, preventing critical analysis of false information.}
We identify the following corruption attacks:
     
     \begin{itemize}
     \itemsep0.5em 
     \item \emph{Spoofing}: Create or modify data in a way that deceives the recipient or system into believing that the data is authentic or unaltered. 
     Two forms of spoofed data are:

         \begin{itemize}[leftmargin=3mm]
         \itemsep0em
         \item \emph{Software Telemetry}: Alter or fabricate telemetry data from software. Attackers create or manipulate telemetry messages that convey a normally functioning application. Further, attackers may spoof telemetry messages at the system level, affecting multiple applications or impacting critical systems \cite{chandio2024stealthy}. 
         
         \item \emph{Hardware Telemetry}: Alter or fabricate telemetry data from hardware sensors. Attackers can generate false sensor readings. Alternatively, attackers can manipulate input data from \MR headsets or peripherals, such as controllers, enacting undesired actions or preventing users from performing desired tasks \cite{tu2018injected, chandio2024stealthy}. 
         \end{itemize}
     
     \item \emph{False-Flag Operations}: Disguise the source of an attack in order to blame another party. 
     \end{itemize} 

     \subsubsection{Subversion} 
     Subversion attacks covertly manipulate a system or its information streams, resulting in falsified and fabricated interpretations by the user. Subversion often employs covert tactics, such as corruption attacks, which weaken trust or disrupt normal operations. 
     \addition{We suspect that the immersiveness of MR can aid false interpretations as users unknowingly engage with deceptive information through repeated interactions, which can correspondingly build trust in deceptive elements.}
     We identify the following subversion attacks:
     \begin{itemize}
     \itemsep0em

     \item \emph{Bias Attacks}: Deliberate manipulation of data or decision-making processes to systematically introduce bias or prejudice toward a specific concept or outcome.
     \item \emph{Disinformation}: Spread false information to deceive and cause harm~\cite{guess2020misinformation}.
     \item  \emph{Lure}: Entice users to engage with (harmful) content.
     \item \emph{Propaganda}: Manipulate perceptions, influence narratives, and garner support for a specific cause or element.
     \item \emph{Gaslighting}: Erode trust and confidence, making it difficult for victims to distinguish truth from deception.
     \end{itemize}


\subsection{Connecting Technical Attacks to Ontology}

\MR deception attacks in our ontology typically rely on technical attacks to facilitate access to \MR systems.
Table~\ref{tab:attacks} characterizes the modalities and deception attacks supported by each technical attack identified in our literature review.
For each technical attack, we identify deception attacks directly mentioned by the authors (\scaledDing{108}) and deception attacks where the technical attack could be deployed but was not specifically mentioned by the authors (\scaledDing{109}).
We found more Channel Attacks (23) mentioned than Processing Attacks (8). 
This is not surprising considering that technical attacks typically target system-level functions which have more impact on the communication channels of \MR headsets than user's cognitive processes.
Still, we see seven attacks that mention Corruption or Subversion, and another eleven that we consider capable of supporting Processing Attacks.

\gap{State-of-the-art \MR technical attacks predominately enable Channel Attacks. More research is needed on technical attacks that facilitate Processing Attacks and how these attacks affect \MR users. }


We identify the sensory modalities affected by an attack and the technical modalities it targets. 
Sensory modalities include visual, auditory, and tactile (e.g., vibrotactile feedback from controllers).
Technical modalities include hardware, software, network, data, and side-channel~\cite{attkan2022cyber}.

\finding{Technical attacks primarily target the visual and software modalities. \MR headsets include displays and processors, making visual and software modalities convenient targets. These attacks particularly focus on overlaying content or replacing signals as opposed to overloading, eroding, or removing signals. The least targeted modalities are tactile and hardware.} 
\section{Information Theory and Deception Attacks}

\label{sec:information-theoretic}

\begin{figure}[ht!]
    \centering
    \includegraphics[width=\columnwidth]{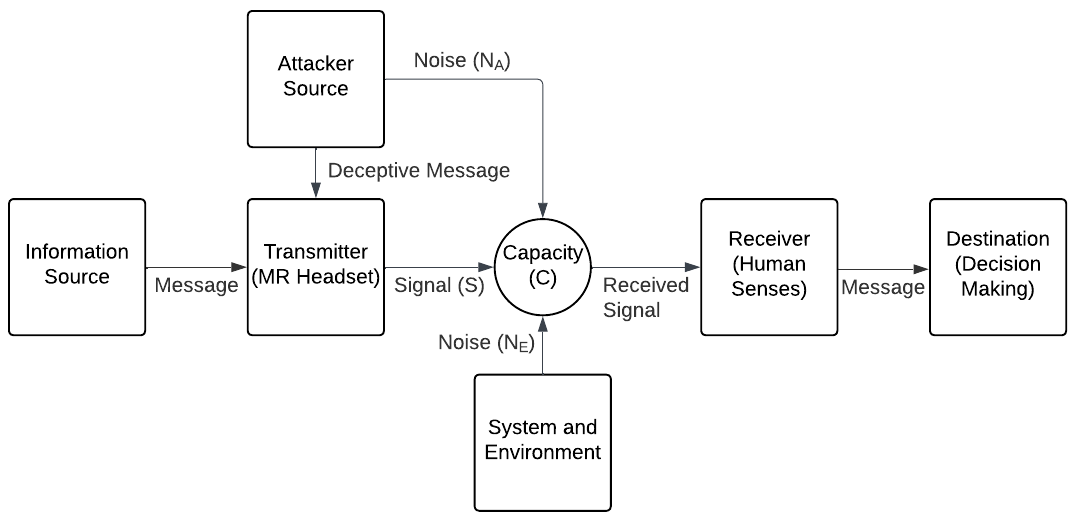}
    \caption{MR Deception Information-Theoretic Model. Messages are transmitted by a \MR headset to a user. Deceptive messages are injected into transmissions. Noise from the attacker or environment affect channel capacity.}
    \label{fig:mr-communication-model}
    \vspace{-1ex}
\end{figure}

\addition{While our ontology categorizes \MR deception attacks, it does not explore the effects of these attacks.} To address \textbf{RQ2}, we use Kopp~et al.'s framework \cite{kopp:2018}, which connects Borden-Kopp's deception model \cite{brumley:2012} and Shannon's communication model \cite{shannon1948mathematical}, to derive an information-theoretic model of \MR deception attacks. 
Shannon's communication model describes how information is transferred from a source to a destination as a message.
The message is sent as a signal through a transmitter to a receiver.
During transmission, the message is affected by noise, which combines with the signal.
In our model, the transmitter is an \MR headset, which acquires information from a source (e.g., an application, sensor, web service), and transmits that information in visual, auditory, or tactile forms to a user (destination) via displays, speakers, and controller vibrations (Figure~\ref{fig:mr-communication-model}).
\MR deception attacks affect the capacity of information transmission by introducing noise to degrade messages, denying access to information, or inserting fake information into messages.



\subsection{Channel Capacity}
According to Shannon's channel capacity theorem \cite{shannon1948mathematical}, the capacity of a channel to transmit information depends upon several factors, including the magnitude of the signal used to encode symbols, the level of interfering noise present in the channel, and the bandwidth of the channel.
\begin{equation} \label{eq:channelcapcity}
    C = W \log_2\left(1 + \frac{S}{N_A + N_E}\right)
\end{equation}
Channel capacity \({C}\) represents the maximum amount of information that can be effectively transmitted from a source to a destination in bits per second (Equation~\ref{eq:channelcapcity}).  Bandwidth \({W}\) refers to the information transfer rate of the communication channel in hertz. As $W$ decreases, channel capacity correspondingly decreases through a linear relationship.

For \MR headsets, information is transmitted through a headset to a human user. 
Thus, channel capacity determines how much visual, auditory, and tactile information can be transmitted. 
Signal \({S}\) is the virtual content transmitted from the headset through displays, speakers, and vibrotactile motors. 
Noise \({N}\) is categorized into two types: \({N_A}\) which represents noise from an attacker source, and \({N_E}\), which represents noise from the real-world environment as well as noise that comes from the system itself, such as rendering stutters or audio glitches. 
\({N_A}\) refers to potential external interference or malicious disruptions.
\({N_E}\) encompasses both ambient disturbances from the surrounding environment and internal system issues that can affect the \MR experience. 
Both these sources of noise have a negative effect the channel capacity. 

\subsection{Channel Attacks}
Channel attacks target channel capacity through reducing bandwidth, manipulating the signal, or introducing noise.
Denial attacks involve an adversary's intention to significantly reduce access to the signal by primarily manipulating bandwidth. The channel capacity \({C}\) tends to zero as the bandwidth \({W}\) tends to zero. 
By shutting down the device, the attacker completely blocks the signal output, and bandwidth ($W$) reduces to zero. 
Attackers can occlude task-specific information with other content, effectively reducing bandwidth and interfering with task performance. 
A Removal attack selectively removes information from the signal, reducing bandwidth as less information is transmitted per a second. 

In Overt Degradation attacks, the adversary can introduce substantial levels of noise into the channel, decreasing the \SNR. As the \SNR tends towards zero, channel capacity \({C}\) decreases and eventually reaches zero.
In this case, the user is bombarded with excessive noise, making it impossible to distinguish between the intended content and the attacker's noise. 
An example of this attack is sensory overload, where an attacker overwhelms the user by emitting excessive sensory stimuli through the \MR headset, resulting in disorientation and discomfort.

In Covert Degradation attacks, an adversary can reduce the signal strength, which results in a decrease in the \SNR. As the signal tends toward zero, \SNR also tends toward zero, decreasing \({C}\) towards zero as well. In \MR headsets, these attacks can involve subtle manipulation of sensory cues presented to a user. Subtle boundary manipulation and subtle dimension manipulation are examples of these attacks. Through subtle manipulation of boundaries or the sizes of virtual objects, the attacker can deceive the user into thinking they are not moving~\cite{Casey_2021} or make it harder to interact with virtual objects.

\subsection{Processing Attacks}
Processing attacks manipulate cognition through deceptive methods that mimic the \MR system.
We use Vitanyi's model~\cite{Vitanyi} to formalize how deceptive information and messages created by an attacker, \({X}\), differ from actual information and messages created by an \MR system, \({Y}\):

\begin{equation}
D(X, Y) = \frac{K(XY) - \min(K(X), K(Y))}{\max(K(X), K(Y))}
\end{equation}

\begin{equation}
M(X, Y) = 1 - D(X, Y)
\end{equation}
where \({D}\) represents the measure of difference, \({M}\) represents the measure of similarity or mimicry, and \({K}\) is the editing function applied to \({X}\) and \({Y}\).

Corruption attacks involve altering data during transmission. Vitanyi's difference measure \({D(X, Y)}\) quantifies the degree of alteration between the original message \({X}\) and the corrupted message \({Y}\).  
In \MR, corruption attacks might involve unauthorized changes to visual information, such as application and system messages, as well as sensory information, including camera, geolocation, and battery status (Figure~\ref{fig:attacks-mindmap}).
Subversion attacks, on the other hand, involve  manipulating how users interpret information within an \MR system. These attacks require repeated corruption or covert degradation attacks to reduce user's trust and understanding.
Thus, $M$ must remain close to $1$ as the user has a greater chance of detecting deceptions through repeated exposure.

\gap{While Vitanyi's model formalizes mimicry, we lack models that effectively describe how processing attacks impact human behavior. Specialized domains, such as formal methods in human-computer interaction, could offer valuable insights.}

\section{Decision-Making and Deception Attacks}
\label{sec:decision-making}

\begin{figure*}[ht!]
\centering
  \includegraphics[width=0.95\textwidth]{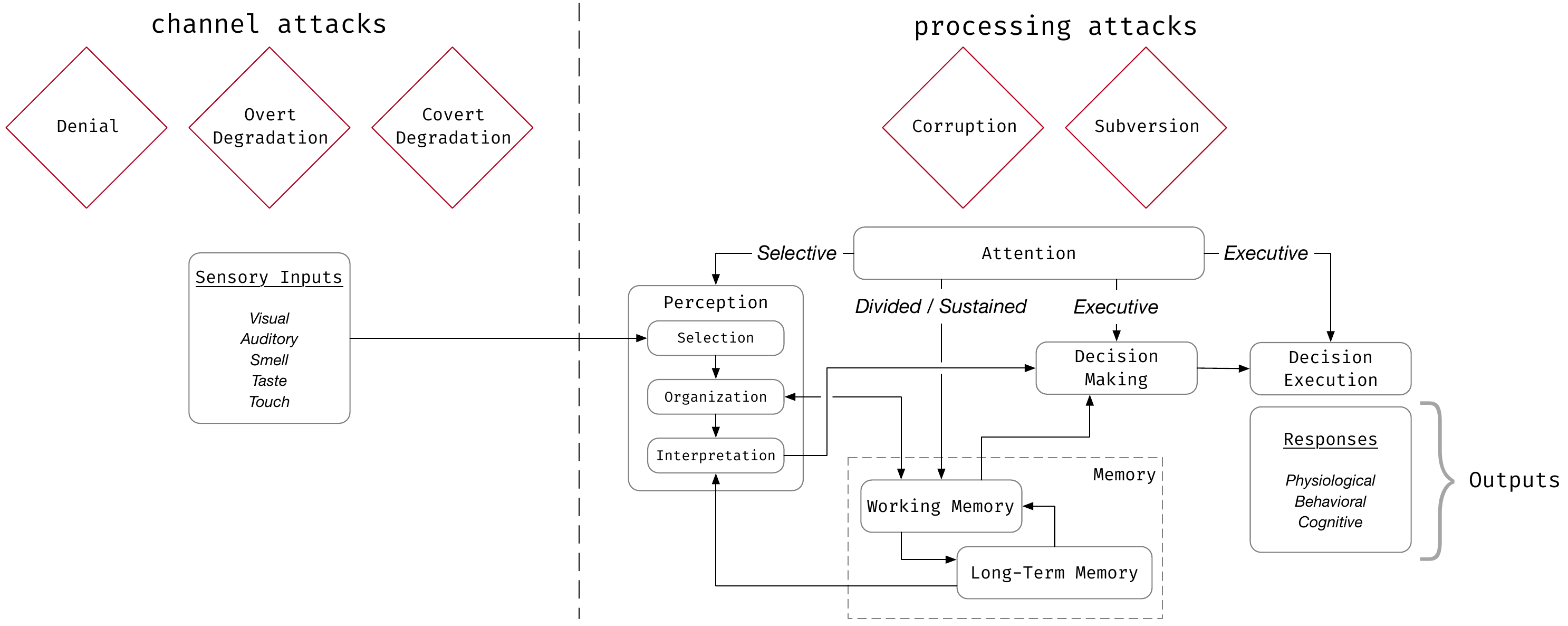}
  \caption{\MR Deception Decision-Making Model. External stimuli (left) are input to cognitive processes (right). Stimuli are first processed by perception. 
  Selective attention manages perception on relevant stimuli. Organized stimuli are stored in working memory. Interpreted stimuli are passed to decision-making, where executive attention manages decisions and their execution.} 
  \label{fig:process-model}
 \vspace{-1ex}
\end{figure*}

\addition{Beyond effects on information channels, we seek to model how \MR deception attacks impact human cognition.} 
To address \textbf{RQ3}, which concerns the interactions between decision-making and \MR deception attacks, \addition{we} conduct a thorough review of the cognition literature and develop a comprehensive decision-making model that outlines the stages of decision-making susceptible to these attacks.
Figure~\ref{fig:process-model} shows our \MR Decision-Making Model. The model provides an overview of how sensory input is cognitively processed by a user to make decisions and where the different types of attacks affect decision making.
Our decision-making model comprises of seven components: Sensory Inputs, Attention, Perception, Memory, Decision-Making, Decision Execution, and Responses.

\begin{itemize}
\itemsep0em
    \item \emph{Sensory Inputs}: Visual, Auditory, Smell, Taste, and Touch are the five different types of sensory inputs that can be affected by deception attacks.
    \item \emph{Attention}: Initial stage where sensory information is gathered. Provides a gateway to perception.
    \item \emph{Perception}: Sensory information gathered is processed and understood.
    \item \emph{Memory}: Processed information is stored in working or long-term memory for future use and retrieval.
    \item \emph{Decision-Making}: Determining a particular course of action predicated on perception.
    \item \emph{Decision Execution}: Decisions are executed.
    \item \emph{Responses}: Physiological, behavioral, or cognitive responses of executed decisions. 
\end{itemize}

\subsection{Perception}
Perception refers to the cognitive process through which one comprehends sensory stimuli~\cite{qiong2017brief}.
Wang et al.~\cite{wang2007cognitive} define perception as ``a set of internal sensational cognitive processes of the brain at the subconscious cognitive function layer that detects, relates, interprets, and searches internal cognitive information in the mind.'' 
Perception is either active or passive.
Active perception involves the intentional direction of attention towards environmental stimuli to extract information \cite{gibson2014ecological}. 
In contrast, passive perception occurs without deliberate effort; sensory information is received as presented \cite{Rock1983-ROCTLO}.


Perception involves three stages~\cite{qiong2017brief}:
\begin{itemize}
\itemsep0em 
    \item \emph{Selection:} Filter and select environmental stimuli from meaningful experiences.
    \item \emph{Organization:} Structure and categorize the selected information, creating coherent and stable perceptions through grouping by proximity and similarity.
    \item \emph{Interpretation:} Assign meaning to organized stimuli, with individuals' cultural or experiential backgrounds leading to different understandings of the same stimuli.
\end{itemize}

In each stage of perception, \MR deception attacks can target specific vulnerabilities. 
During selection, attacks can cause sensory overload or misdirect focus on irrelevant stimuli. In the organization step, attacks could involve boundary or dimension manipulation, affecting how stimulus are structured and grouped due to changes in proximity or scale. Propaganda or spoofing attacks can target interpretation, affecting the meaning assigned to stimuli that may seem wrong but is coming from a trusted source (e.g., the system or a collaborator). These potential attacks highlight the importance of the accuracy and reliability of perception in \MR systems.

In addition to the conscious components of perception, subliminal inputs play an important role in how individuals interact with and understand \MR environments.
Cetnarski et al.\ \cite{Cetnarski2014Subliminal} show that subliminal stimuli—information presented below the threshold of conscious awareness—can significantly influence decision-making processes in \MR. This underscores the need to understand these subtle interactions that occur at the subconscious level of perception.
\subsection{Attention} 

James~\cite{James1890} described attention as the cognitive process by which the mind selectively concentrates on a singular element from a variety of possible stimuli or thoughts, emphasizing its essential function in creating our conscious perception. 
The seminal work of Posner \cite{posner1980orienting} 
introduced a framework for understanding the neural bases of attention and its various components and extending James's initial descriptions into a more nuanced understanding of the brain's attentional mechanisms. Building upon these early foundations, attention classification has expanded to include four types: 


\begin{itemize}
    \itemsep0em
    \item \emph{Selective:} Focusing on relevant information while suppressing irrelevant information \cite{stevens2012role, murphy2016twenty}.
    
    \item \emph{Divided:} Capacity to allocate cognitive resources to multiple stimuli simultaneously, enabling individuals to engage in concurrent activities~\cite{spelke1976skills}. 
    Attended stimuli are from the same sensory modality or different ones~\cite{Herbranson2017}.

    \item \emph{Sustained:} Readiness to perceive and respond to stimuli over prolonged periods, often without conscious awareness of this vigilance~\cite{mackworth1948breakdown}.
    \item \emph{Executive:} Regulates cognitive and emotional responses through management of other cognitive processes~\cite{posner1990attention}. Aids orchestration of thought and emotion in alignment with goals and the dynamic demands of the environment.
\end{itemize}

Channel attacks primarily target Selective and Sustained attention. 
They manipulate the sensory channels through which users receive information, affecting their ability to focus on relevant stimuli or maintain attention over time.
Selective attention is exploited by degrading the sensory inputs, making it harder for users to distinguish between relevant and irrelevant stimuli. As mentioned in Section \ref{sec:decattacks}, this happens in overt and covert degradation attacks. These attacks may also limit sustained attention by making it more challenging for the user to maintain their focus over time, particularly when the quality of sensory inputs fluctuates or declines, resulting in increased cognitive load. 
Denial attacks block access to certain stimuli or information channels, disrupting selective attention. 

Processing attacks primarily affect Divided and Executive attention by overloading the cognitive processing capabilities or by requiring constant adjustments to unexpected system behaviors.
Corruption attacks 
can directly impact the users' selective and executive attention by altering the information presented within \MR environment and also exploiting perceptual biases.
Subversion attacks could challenge executive attention by forcing users to constantly adapt to unexpected system responses, requiring continuous updating of working memory. They also can target divided attention by interrupting the flow of tasks or actions within an \MR environment, which compels users to divide their attention between correcting system errors and accomplishing their original goals.

\finding{Perception and attention are the primary targets for \MR deception attacks. Channel attacks target selection mechanisms by degrading or denying stimuli. Processing attacks target interpretation and execution by corrupting beliefs or subverting interpretations.}
\vspace{-1ex}


\definecolor{lowcolor}{HTML}{bcabcd} 
\definecolor{midcolor}{HTML}{8a6ca8} 
\definecolor{highcolor}{HTML}{582D83}
\definecolor{none}{HTML}{ffffff}


\newcommand{\Low}{%
\begin{tikzpicture}
\fill[fill=lowcolor] (0,0) rectangle (0.33,0.2);
\fill[fill=none] (0.33,0) rectangle (0.66,0.2);
\fill[fill=none] (0.66,0) rectangle (1,0.2);
\end{tikzpicture}%
}

\newcommand{\Medium}{%
\begin{tikzpicture}
\fill[fill=midcolor] (0,0) rectangle (0.33,0.2);
\fill[fill=midcolor] (0.33,0) rectangle (0.66,0.2);
\fill[fill=none] (0.66,0) rectangle (1,0.2);
\end{tikzpicture}%
}

\newcommand{\High}{%
\begin{tikzpicture}
\fill[fill=highcolor] (0,0) rectangle (1,0.2); 
\end{tikzpicture}%
}

\newcommand{\LowMedium}{%
\begin{tikzpicture}
\fill[fill=lowcolor] (0,0) rectangle (0.33,0.2);
\fill[fill=midcolor] (0.33,0) rectangle (0.66,0.2);
\fill[fill=none] (0.66,0) rectangle (1,0.2);
\end{tikzpicture}%
}

\newcommand{\MediumHigh}{%
\begin{tikzpicture}
\fill[fill=none] (0,0) rectangle (0.33,0.2);
\fill[fill=midcolor] (0.33,0) rectangle (0.66,0.2);
\fill[fill=highcolor] (0.66,0) rectangle (1,0.2);
\end{tikzpicture}%
}

\newcommand{\LowHigh}{%
\begin{tikzpicture}
\fill[fill=lowcolor] (0,0) rectangle (0.33,0.2);
\fill[fill=midcolor] (0.33,0) rectangle (0.66,0.2);
\fill[fill=highcolor] (0.66,0) rectangle (1,0.2);
\end{tikzpicture}%
}

\begin{table*}[ht!]
\caption{\MR attacks from our ontology are assessed according to the Information-Theoretic Model and Decision-Making Model.}


\small
\centering
\renewcommand{\arraystretch}{1.15}
\setlength{\tabcolsep}{1pt}

\begin{tabular}{ccr|cccc|cccccccc|c|}
\cline{4-16}
  & &  & 
  \multicolumn{4}{c|}{Information-} & 
  \multicolumn{9}{c|}{} \\   

  & &  & 
  \multicolumn{4}{c|}{\multirow{-1}{*}{Theoretic Model}} & 
  \multicolumn{9}{c|}{\multirow{-2}{*}{Decision-Making Model}} \\ \cline{4-16} 

  & &  & 
  \multicolumn{3}{c|}{$C$} & & 
  \multicolumn{4}{c|}{Perception} & 
  \multicolumn{4}{c|}{Attention} & 
  \multirow{2}{*}{Mem.} \\ \cline{4-6} \cline{8-15} 

  &
   &
  \multicolumn{1}{r|}{\multirow{-2}{*}{ \backslashbox{Attacks}{Models}}} &
  \multicolumn{1}{c|}{\hspace*{1mm}$W$\hspace*{1mm}} &
  \multicolumn{1}{c|}{\hspace*{1mm}$S$\hspace*{1mm}} &
  \multicolumn{1}{c|}{\hspace*{1mm}$N$\hspace*{1mm}} &
  \multirow{-2}{*}{$M$} &
  \multicolumn{1}{c|}{\hspace*{0.5mm}A/P\hspace*{0.5mm}} &
  \multicolumn{1}{c|}{$Sel$} &
  \multicolumn{1}{c|}{$Org$} &
  \multicolumn{1}{c|}{$Int$} &
  \multicolumn{1}{c|}{$Foc$} &
  \multicolumn{1}{c|}{$Div$} &
  \multicolumn{1}{c|}{$Sus$} &
  \multicolumn{1}{c|}{$Exe$} &
  \multicolumn{1}{c|}{} \\ \hlinewd{1.5pt}

  \multicolumn{1}{|c|}{} & \multicolumn{1}{c|}{}
   &
  Sensory Overload~ &
  \multicolumn{1}{c|}{} &
  \multicolumn{1}{c|}{} &
  \multicolumn{1}{c|}{\ding{51}} &
   &
  \multicolumn{1}{c|}{A} &
  \multicolumn{1}{c|}{\High} &
  \multicolumn{1}{c|}{\High} &
  \multicolumn{1}{c|}{\High} &
  \multicolumn{1}{c|}{\High} &
  \multicolumn{1}{c|}{\High} &
  \multicolumn{1}{c|}{\High} &
  \multicolumn{1}{c|}{\High} &
  \multicolumn{1}{c|}{W} \\ 
   \multicolumn{1}{|c|}{} & \multicolumn{1}{c|}{}
   &
  Momentary Misdirection~ & 
  \multicolumn{1}{c|}{} &
  \multicolumn{1}{c|}{} &
  \multicolumn{1}{c|}{\ding{51}} &
   &
  \multicolumn{1}{c|}{A} &
  \multicolumn{1}{c|}{\LowHigh} &
  \multicolumn{1}{c|}{\LowHigh} &
  \multicolumn{1}{c|}{\LowHigh} &
  \multicolumn{1}{c|}{\High} &
  \multicolumn{1}{c|}{\LowHigh} &
  \multicolumn{1}{c|}{\LowHigh} &
  \multicolumn{1}{c|}{\LowHigh} &
  \multicolumn{1}{c|}{W} \\ 
  \multicolumn{1}{|c|}{} &
  \multicolumn{1}{c|}{\multirow{-3}{*}{Overt} }&
  Signal Replacement~ &
  \multicolumn{1}{c|}{} &
  \multicolumn{1}{c|}{\ding{51}} &
  \multicolumn{1}{c|}{} &
   &
  \multicolumn{1}{c|}{A} &
  \multicolumn{1}{c|}{\High} &
  \multicolumn{1}{c|}{\High} &
  \multicolumn{1}{c|}{\High} &
  \multicolumn{1}{c|}{\High} &
  \multicolumn{1}{c|}{\High} &
  \multicolumn{1}{c|}{\High} &
  \multicolumn{1}{c|}{\High} &
  \multicolumn{1}{c|}{W} \\  
 \multicolumn{1}{|c|}{} & 
  \multicolumn{1}{c|}{\multirow{-3}{*}{Degradation}} & 
  Quality Erosion~ &
  \multicolumn{1}{c|}{\ding{51}} &
  \multicolumn{1}{c|}{} &
  \multicolumn{1}{c|}{\ding{51}} &
   &
  \multicolumn{1}{c|}{A} &
  \multicolumn{1}{c|}{\LowHigh} &
  \multicolumn{1}{c|}{\LowHigh} &
  \multicolumn{1}{c|}{\LowHigh} &
  \multicolumn{1}{c|}{\LowHigh} &
  \multicolumn{1}{c|}{\LowHigh} &
  \multicolumn{1}{c|}{\LowHigh} &
  \multicolumn{1}{c|}{\LowHigh} &
  \multicolumn{1}{c|}{W} \\ \cline{2-16} 
   \multicolumn{1}{|c|}{} & \multicolumn{1}{c|}{} &
  Subtle Boundary Manipulation~  &
  \multicolumn{1}{c|}{} &
  \multicolumn{1}{c|}{\ding{51}} &
  \multicolumn{1}{c|}{} &
   &
  \multicolumn{1}{c|}{P} &
  \multicolumn{1}{c|}{\Low} &
  \multicolumn{1}{c|}{\Low} &
  \multicolumn{1}{c|}{\LowHigh} &
  \multicolumn{1}{c|}{\Low} &
  \multicolumn{1}{c|}{\Low} &
  \multicolumn{1}{c|}{\Low} &
  \multicolumn{1}{c|}{\LowMedium} &
  \multicolumn{1}{c|}{W} \\ 
 \multicolumn{1}{|c|}{} &
  \multicolumn{1}{c|}{\multirow{-2}{*}{Covert}} &
  Subtle Dimension Manipulation~  &
  \multicolumn{1}{c|}{} &
  \multicolumn{1}{c|}{\ding{51}} &
  \multicolumn{1}{c|}{} &
   &
  \multicolumn{1}{c|}{P} &
  \multicolumn{1}{c|}{\Low} &
  \multicolumn{1}{c|}{\LowMedium} &
  \multicolumn{1}{c|}{\LowHigh} &
  \multicolumn{1}{c|}{\Low} &
  \multicolumn{1}{c|}{\Low} &
  \multicolumn{1}{c|}{\Low} &
  \multicolumn{1}{c|}{\LowMedium} &
  \multicolumn{1}{c|}{W} \\
 \multicolumn{1}{|c|}{} &
  \multicolumn{1}{c|}{\multirow{-2}{*}{Degradation}} &
  Physical Movement Manipulation~ &
  \multicolumn{1}{c|}{} &
  \multicolumn{1}{c|}{\ding{51}} &
  \multicolumn{1}{c|}{} &
   &
  \multicolumn{1}{c|}{P} &
  \multicolumn{1}{c|}{\Low} &
  \multicolumn{1}{c|}{\Low} &
  \multicolumn{1}{c|}{\LowHigh} &
  \multicolumn{1}{c|}{\Low} &
  \multicolumn{1}{c|}{\Low} &
  \multicolumn{1}{c|}{\Low} &
  \multicolumn{1}{c|}{\LowHigh} &
  \multicolumn{1}{c|}{W} \\ \cline{2-16} 
 \multicolumn{1}{|c|}{} & \multicolumn{1}{c|}{}
   &
  Shutdown~ &
  \multicolumn{1}{c|}{\ding{51}} &
  \multicolumn{1}{c|}{} &
  \multicolumn{1}{c|}{} &
   &
  \multicolumn{1}{c|}{A} &
  \multicolumn{1}{c|}{\High} &
  \multicolumn{1}{c|}{\High} &
  \multicolumn{1}{c|}{\High} &
  \multicolumn{1}{c|}{\High} &
  \multicolumn{1}{c|}{\High} &
  \multicolumn{1}{c|}{\High} &
  \multicolumn{1}{c|}{\High} &
  \multicolumn{1}{c|}{W} \\ 
  \multicolumn{1}{|c|}{} & \multicolumn{1}{c|}{}
   &
  Overlay~ &
  \multicolumn{1}{c|}{\ding{51}} &
  \multicolumn{1}{c|}{} &
  \multicolumn{1}{c|}{\ding{51}} &
   &
  \multicolumn{1}{c|}{A} &
  \multicolumn{1}{c|}{\LowHigh} &
  \multicolumn{1}{c|}{\LowHigh} &
  \multicolumn{1}{c|}{\LowHigh} &
  \multicolumn{1}{c|}{\LowHigh} &
  \multicolumn{1}{c|}{\LowHigh} &
  \multicolumn{1}{c|}{\LowHigh} &
  \multicolumn{1}{c|}{\LowHigh} &
  \multicolumn{1}{c|}{W} \\  
\multicolumn{1}{|c|}{\multirow{-10}{*}{\rotatebox[origin=c]{90}{Channel Attacks}}} &
  \multicolumn{1}{c|}{\multirow{-3}{*}{Denial}} & 
  Removal~ &
  \multicolumn{1}{c|}{\ding{51}} &
  \multicolumn{1}{c|}{\ding{51}} &
  \multicolumn{1}{c|}{} &
   &
  \multicolumn{1}{c|}{A} &
  \multicolumn{1}{c|}{\LowHigh} &
  \multicolumn{1}{c|}{\LowHigh} &
  \multicolumn{1}{c|}{\LowHigh} &
  \multicolumn{1}{c|}{\LowHigh} &
  \multicolumn{1}{c|}{\LowHigh} &
  \multicolumn{1}{c|}{\LowHigh} &
  \multicolumn{1}{c|}{\LowHigh} &
  \multicolumn{1}{c|}{W} \\ \cline{1-16}
 \multicolumn{1}{|c|}{} &
  \multicolumn{1}{c|}{} &
  Spoofing~ &
  \multicolumn{1}{c|}{} &
  \multicolumn{1}{c|}{} &
  \multicolumn{1}{c|}{} &
  \ding{51} &
  \multicolumn{1}{c|}{P} &
  \multicolumn{1}{c|}{\Low} &
  \multicolumn{1}{c|}{\Low} &
  \multicolumn{1}{c|}{\LowHigh} &
  \multicolumn{1}{c|}{\LowHigh} &
  \multicolumn{1}{c|}{\LowHigh} &
  \multicolumn{1}{c|}{\LowHigh} &
  \multicolumn{1}{c|}{\LowHigh} &
  \multicolumn{1}{c|}{W} \\ 
 \multicolumn{1}{|c|}{} &
  
  \multicolumn{1}{c|}{\multirow{-2}{*}{Corruption}} &
  False-Flag Operations~ &
  \multicolumn{1}{c|}{} &
  \multicolumn{1}{c|}{} &
  \multicolumn{1}{c|}{} &
  \ding{51} &
  \multicolumn{1}{c|}{P} &
\multicolumn{1}{c|}{\Low} &
  \multicolumn{1}{c|}{\Low} &
  \multicolumn{1}{c|}{\LowHigh} &
  \multicolumn{1}{c|}{\Low} &
  \multicolumn{1}{c|}{\Low} &
  \multicolumn{1}{c|}{\Low} &
  \multicolumn{1}{c|}{\Low} &
  \multicolumn{1}{c|}{W/L} \\ \cline{2-16} 
  \multicolumn{1}{|c|}{} &
  \multicolumn{1}{c|}{} &
  Bias Attacks~ &
  \multicolumn{1}{c|}{} &
  \multicolumn{1}{c|}{} &
  \multicolumn{1}{c|}{} &
   &
  \multicolumn{1}{c|}{P} &
  \multicolumn{1}{c|}{\Low} &
  \multicolumn{1}{c|}{\Low} &
  \multicolumn{1}{c|}{\LowHigh} &
  \multicolumn{1}{c|}{\Low} &
  \multicolumn{1}{c|}{\Low} &
  \multicolumn{1}{c|}{\Low} &
  \multicolumn{1}{c|}{\Low} &
  \multicolumn{1}{c|}{W/L} \\  
  \multicolumn{1}{|c|}{} &
  \multicolumn{1}{c|}{} &
  Lure~ &
  \multicolumn{1}{c|}{} &
  \multicolumn{1}{c|}{} &
  \multicolumn{1}{c|}{} &
   &
  \multicolumn{1}{c|}{P} &
  \multicolumn{1}{c|}{\LowHigh} &
  \multicolumn{1}{c|}{\LowHigh} &
  \multicolumn{1}{c|}{\LowHigh} &
  \multicolumn{1}{c|}{\LowHigh} &
  \multicolumn{1}{c|}{\LowHigh} &
  \multicolumn{1}{c|}{\LowHigh} &
  \multicolumn{1}{c|}{\LowHigh} &
  \multicolumn{1}{c|}{W/L} \\  
  \multicolumn{1}{|c|}{} &
  \multicolumn{1}{c|}{} &
  Disinformation~ &
  \multicolumn{1}{c|}{} &
  \multicolumn{1}{c|}{} &
  \multicolumn{1}{c|}{} &
   &
  \multicolumn{1}{c|}{P} &
  \multicolumn{1}{c|}{\Low} &
  \multicolumn{1}{c|}{\Low} &
  \multicolumn{1}{c|}{\LowHigh} &
  \multicolumn{1}{c|}{\Low} &
  \multicolumn{1}{c|}{\Low} &
  \multicolumn{1}{c|}{\Low} &
  \multicolumn{1}{c|}{\Low} &
  \multicolumn{1}{c|}{W/L} \\ 
  \multicolumn{1}{|c|}{} &
  \multicolumn{1}{c|}{} &
  Propaganda~ &
  \multicolumn{1}{c|}{} &
  \multicolumn{1}{c|}{} &
  \multicolumn{1}{c|}{} &
   &
  \multicolumn{1}{c|}{P} &
  \multicolumn{1}{c|}{\Low} &
  \multicolumn{1}{c|}{\Low} &
  \multicolumn{1}{c|}{\LowHigh} &
  \multicolumn{1}{c|}{\Low} &
  \multicolumn{1}{c|}{\Low} &
  \multicolumn{1}{c|}{\Low} &
  \multicolumn{1}{c|}{\Low} &
  \multicolumn{1}{c|}{W/L} \\ 
\multicolumn{1}{|c|}{\multirow{-7}{*}{\rotatebox[origin=c]{90}{Processing Attacks}}} &
  \multicolumn{1}{c|}{\multirow{-5}{*}{Subversion}} &
  Gaslighting~ &
  \multicolumn{1}{c|}{} &
  \multicolumn{1}{c|}{} &
  \multicolumn{1}{c|}{} &
   &
  \multicolumn{1}{c|}{P} &
  \multicolumn{1}{c|}{\Low} &
  \multicolumn{1}{c|}{\Low} &
  \multicolumn{1}{c|}{\LowHigh} &
  \multicolumn{1}{c|}{\Low} &
  \multicolumn{1}{c|}{\Low} &
  \multicolumn{1}{c|}{\Low} &
  \multicolumn{1}{c|}{\LowHigh} &
  \multicolumn{1}{c|}{W/L} \\ \hline
\end{tabular}
\\ \vspace{0.5em}
Low = 
\begin{tikzpicture}
\fill[fill=lowcolor] (0,0) rectangle (0.33,0.2); 
\end{tikzpicture}, 
Low-Medium = \begin{tikzpicture}
\fill[fill=lowcolor] (0,0) rectangle (0.33,0.2);
\fill[fill=midcolor] (0.33,0) rectangle (0.66,0.2);
\end{tikzpicture}, Low-High = \LowHigh, High = \High \protect\\
  \textbf{Information-Theoretic Model}: $C =$ Channel Capacity, $W =$ Bandwidth, $S =$ Signal, $N =$ Noise, $M =$ Mimicry \protect\\ \textbf{Perception}: $A/P =$ Active/Passive, $Sel =$ Selection, $Org =$ Organization, $Int =$ Interpretation \protect\\ \textbf{Attention}: $Foc =$ Selective, $Div =$ Divided, $Sus =$ Sustained, $Exe =$ Executive; \textbf{Memory}: $W =$ Working, $L =$ Long-Term  
\label{tab:framework}
\vspace{-1ex}
\end{table*}

\subsection{Memory}
Working memory and long-term memory are central components of our decision-making model. 
Baddeley~\cite{baddeley1974working,alan1992working,baddeley2007working} derived a multicomponent model of working memory consisting of the visuospatial sketchpad, phonological loop, central executive, and episodic buffer.
The visuospatial sketchpad stores visual and spatial information while the phonological loop stores auditory and verbal information.
The central executive directs attention towards \minoraddition{stored} information \minoraddition{in either one}.
The episodic buffer provides temporary storage of information needed by the central executive with connections to the other three components and long-term memory.
Long-term memory represents a permanent store that receives selected inputs from both the sensory register and working memory~\cite{ATKINSON196889}.


\MR deception attacks affect memory and correspondingly attention.
Downing \cite{downing2000interactions} showed that the content of visuospatial sketchpad can guide selective attention toward matching visual stimuli. 
Through spoofing attacks, adversaries can produce deceptive stimuli that match expected stimuli, leveraging working memory to direct the user's selective attention.
Santangelo and Macaluso~\cite{santangelo2013contribution} identified the critical role of working memory in managing divided attention
when monitoring multiple objects simultaneously. 
Working memory load directly affects the efficiency of the central executive, with increased load impairing attention to multiple stimuli. 
Thus, sensory overload attacks can overwhelm working memory by visualizing too many objects for working memory to maintain.
Unsworth \& Robinson \cite{unsworth2020working} suggested 
that individuals with lower \WMC may struggle more with maintaining consistent attention, leading to performance degradation in tasks requiring prolonged focus. 
Therefore, the impact of \MR deception attacks that target \WMC, such as sensory overload, will vary from person to person.
\section{\MR Deception Analysis Framework (DAF)}
\label{sec:cog-framework}
The culminating, ensemble knowledge that connects our ontology, information-theoretic model, and decision-making model is the \MR \acf{DAF}---an assessment tool for identifying and discussing the multifaceted impact of \MR deception attacks on user cognition (\textbf{RQ4}).


\DAF classifies attacks according to their operational mechanisms, which can be overt or covert, involving Degradation, Denial, Corruption, or Subversion, and the cognitive processes they aim to disrupt. 
We focus on identifying where attacks manipulate \MR communication channels by altering bandwidth ($W$), signal ($S$), noise ($N$), or by employing mimicry ($M$). Additionally, we explore the cognitive effects of each attack, examining the extent to which they can affect perception, attention, and memory.
For perception and attention, we further breakdown analysis into stages of perception---Selection ($Sel$), Organization ($Org$), and Interpretation ($Int$)---and types of attention---Selective ($Sel$), Divided ($Div$), Sustained ($Sus$), and Executive ($Exe$).

Table~\ref{tab:framework} presents our general analysis of the different categories of attacks identified in our ontology.
Overt Degradation and Denial attacks strongly affect both perception and attention.
Covert Degradation, Corruption, and Subversion attacks primarily target the Interpretation stage of perception.
These attacks typically require remaining hidden from the user.
Thus, any effects on attention or early stages of perception are likely too revealing.

\finding{The interpretation stage of perception is a primary target of \MR deception attacks. Deceptions seek to cultivate false beliefs, formed initially by interpretations of perceived stimuli.}



For assessing the degree to which attacks affect stages of perception, we derived the following questions. Answers are either Low, Medium, High, or a combination of the three. \addition{De Meyer et al.~\cite{de2019delphi} state that a three-point scale provides a practical balance between simplicity and reliability. It minimizes measurement error and ensures clarity in response, which can be important for consensus building in Delphi procedures.} 

\begin{itemize}
\itemsep0em
    \item \emph{Selection:} To what degree does the attack make it difficult to attend to or ignore task-related sensory stimuli from the physical or virtual environments during a decision-making task? 
    \item \emph{Organization:} To what degree does the attack make it difficult to group task-related sensory stimuli, such as by proximity or similarity, for a decision-making task?
    \item \emph{Interpretation:} To what degree does the attack make it difficult to accurately assign meaning to organized, task-related stimuli and correctly interpret patterns and relationships within virtual and physical environments when deriving understanding, making decisions, and taking action in a decision-making task?
\end{itemize}    

For assessing the degree to which attacks affect types of attention, we derived the following questions. Answers are either Low, Medium, High, or a combination of the three.
\begin{itemize}
\itemsep0em
    \item \emph{Selective:} To what degree does the attack make it difficult to focus attention on relevant physical and virtual objects for a decision-making task in MR?
    \item \emph{Divided:} To what degree does the attack make it difficult to switch between concurrent tasks rapidly while maintaining situational awareness in both the virtual and physical environments?
    \item \emph{Sustained:} To what degree does the attack make it difficult to continuously scan and interpret information presented in the mixed reality environment, making timely decisions and adjustments?
    \item \emph{Executive:} To what degree does the attack make it difficult to manage attentional resources effectively to interact with virtual elements while remaining aware of and responsive to the physical environment while performing a decision-making task?    
\end{itemize}

\DAF provides a systematic approach to evaluate threats posed by \minoraddition{\MR} deception attacks. We posit that such analysis is pivotal for developing more resilient \MR systems and training programs that can mitigate the impacts of deceptive threats. 

\finding{\DAF is a tool for defining experimental research on \MR deception attacks. We posit that it can be used to explore future attacks and may be extended for deception analysis beyond \MR research.}

\gap{We need empirical findings to validate and precisely model the impact of \MR deception attacks on cognitive processes and information channels.}

\addition{
\section{Discussion}
\label{sec:discussion}
\DAF provides a systematic method to classify and analyze \MR deception attacks. 
While we focus on \MR headsets, \DAF is applicable to other forms of \MR and even other areas of human-computer interaction (HCI).
Kopp et al.'s information-theoretic framework ~\cite{kopp:2018} applied the Borden-Kopp model of deception to news media.
We have broadened its use to \MR deception attacks.
Future work should extend the scope to other areas of HCI that involve information processing and decision-making.
Our information-theoretic model and decision-making model are not tied to specific technologies or attacks, but rather provide generalizable models for studying the effects of deception in computing.
To enhance \DAF, future work should validate it empirically, expand its applicability to diverse contexts, incorporate individual cognitive factors, and refine models for processing attacks.

Researchers and practitioners can use \DAF to assess the security threat of \MR deception attacks.
For example, we can assign values of 1 to 3 for Low to High ratings, respectively.
Then, we can sum the values to identify which attacks pose the highest threat to perception and attention.
Further, \DAF can help develop deception detection and prevention approaches. 
For example, we can compare differences between rendered frames to see how the signal is changing.
High volatility in changes may indicate overt degradation attacks, particularly if we can identify noise based on differences between expected and actual frames.
More subtle changes that are spatial located in unexpected areas may indicate covert degradation attacks.
Using eye-tracking sensors on these headsets, we can derive models of attention that can help identify when different types of attention are being employed or disrupted.


}

\addition{
\textbf{Limitations:} This SoK synthesizes existing knowledge towards developing a field of study around \MR deception. 
It is theoretical in nature and would benefit from further empirical validation.
Controlled experiments involving \MR deception attacks are essential for refining the framework and assessing its relevance to diverse scenarios. 
Furthermore, \DAF does not fully account for cognitive diversity among users. 
Individual differences in cognitive capacity, attention, and susceptibility to deception are critical factors that could influence the effectiveness of both attacks and countermeasures. 
}

\section{Conclusion}
\label{sec:conclusion}

%
\MR technologies provide a wide range of opportunities while raising significant cybersecurity challenges. 
\MR systems influence how we perceive physical and virtual environments, making them particularly susceptible to deception attacks.
This multidisciplinary SoK brings together diverse knowledge to provide a systematic way for categorizing and analyzing \MR deceptive attacks and their effects. 
It serves as a foundation and guide for future research.
Our examination indicates that while there is a growing interest in \MR security and a number of technical attacks, there is a lack of comprehensive research regarding deception attacks, particularly with regards to information communication and human cognition. 
Future work should investigate empirical studies of how \MR deception attacks affect cognitive processes.
Such studies can inform new techniques for detecting and mitigating threats from \MR deception attacks.
\addition{We envision \DAF as a generalizable framework; however, specialized domains may possess nuances not fully captured by \DAF.
While we expect our information-theoretic approach to remain valid across \MR technologies, where information is still transmitted and processed,
it may be necessary to extend \DAF through new metrics and additional models.
We look forward to seeing how researchers can leverage \DAF in studies of deception.}





\section*{Acknowledgments}
\label{sec:ack}

This work was supported by the \DARPA, grant number HR00112320030.
We appreciate the shepherd and anonymous reviewers for their insightful and valuable feedback.
\section*{Ethics Considerations}
\label{sec:ethics}

Conducting research on deception requires significant ethical considerations.
At the forefront is mitigating risks to human participants and end users.
When deceiving participants, it is necessary to ensure that benefits of the research far outweigh any potential risks to participants.
Typically, research institutions have an IRB to enforce participant protections from unnecessary harm during human-subjects research.
For \MR research, harm can take many forms including physical, cognitive, technological, and social.
As \MR headsets affect how users perceive the physical world, deception attacks pose significant physical risk.
Precautions must be taken to mitigate risks by screening out participants that may have adverse reactions to perceptual manipulations.
Further, researchers should provide safe environments where participants cannot harm themselves by colliding with objects or falling down.
Researchers should also consider how deceptive information may impact participants trust and understanding of \MR systems.
Studies require effective debriefing that helps the participant understand how they were deceived, what elements were deceptive, and how to evaluate potential deceptions.
While this SoK synthesizes knowledge from diverse domains, it does not directly involve human-subjects research or development of interactive systems.
However, we do provide a framework for exploring cognitive and technological harm of deception attacks in \MR.

\addition{
\section*{Open Science}
The primary artifact for this SoK is a comprehensive list of articles analyzed during development of the ontology and corresponding \MR Deception Analysis Framework. 
This list includes articles cited in this work as well as others that are not cited. A link to this list can be found at: \url{https://doi.org/10.5281/zenodo.14732979}. No other research artifacts, besides diagrams and tables presented in this paper, resulted from this research.
}

\nocite{*}
{\footnotesize \bibliographystyle{plain}
\bibliography{main}}



\end{document}